\pdfoutput=1
\documentclass[fleqn,usenatbib]{mnras}

\usepackage{savesym}
\usepackage{amsmath}
\savesymbol{iint}
\usepackage{txfonts}
\restoresymbol{TXF}{iint}
\usepackage[T1]{fontenc}
\usepackage{ae,aecompl}

\usepackage{graphicx}	
\usepackage{amssymb}	
\usepackage{gensymb}

\title[Jet or Shock Breakout? GRB 060218]{Jet or Shock Breakout? The Low-Luminosity GRB 060218}

\author[Irwin \& Chevalier]{
Christopher M. Irwin,$^{1}$
Roger A. Chevalier$^{1}$
\\
$^{1}$Department of Astronomy, University of Virginia, Charlottesville, VA, 22904}

\date{Accepted XXX. Received YYY; in original form ZZZ}

\pubyear{2015}

\begin{document}
\label{firstpage}
\pagerange{\pageref{firstpage}--\pageref{lastpage}}
\maketitle

\begin{abstract}
We consider a model for the low-luminosity gamma-ray burst GRB 060218 that plausibly accounts for multiwavelength observations to day 20.  The model components are: (1) a long-lived ($t_j \sim 3000$ s) central engine and accompanying low-luminosity ($L_j \sim 10^{47}$\,erg\,s$^{-1}$), semirelativistic ($\gamma \sim 10$) jet; (2) a low-mass ($\sim 4 \times 10^{-3} M_\odot$) envelope surrounding the progenitor star; and (3) a modest amount of dust ($A_V \sim 0.1$\,mag) in the interstellar environment.  Blackbody emission from the transparency radius in a low-power jet outflow can fit the prompt thermal X-ray emission, and the nonthermal X-rays and $\gamma$-rays may be produced via Compton scattering of thermal photons from hot leptons in the jet interior or the external shocks.  The later mildly relativistic phase of this outflow can produce the radio emission via synchrotron radiation from the forward shock.  Meanwhile, interaction of the associated SN 2006aj with a circumstellar envelope extending to $\sim 10^{13}$ cm can explain the early optical emission.  The X-ray afterglow can be interpreted as a light echo of the prompt emission from dust at $\sim 30$ pc.  Our model is a plausible alternative to that of Nakar, who recently proposed shock breakout of a jet smothered by an extended envelope as the source of prompt emission.  Both our results and Nakar's suggest that bursts such as GRB 060218 may originate from unusual progenitors with extended circumstellar envelopes, and that a jet is necessary to decouple the prompt emission from the supernova.  
\end{abstract}


%---------------------------------------------------------------------------------------------------------------------------
%********************************************SECTION 1************************************************
%---------------------------------------------------------------------------------------------------------------------------

\section{Introduction}
\label{intro}

Low-luminosity gamma-ray bursts (LLGRBs) are a subclass of long-duration gamma-ray bursts (GRB) that, although rarely detected and not yet well understood, have the potential to shed light on more commonly observed cosmological bursts.  Though the uncertainty is large, estimated volumetric rates indicate that LLGRBs occur some $10 - 100$ times more often than typical GRBs \citep{soderberg06}, making them a compelling population for further study.  In addition, LLGRBs take place nearby, so the associated supernovae (SNe) can be detected easily and studied in detail, placing constraints on energetics and circumstellar environment and giving clues about the SN-GRB connection.   Phenomena like central engine activity, jet-star and jet-wind interactions, and the transition from beamed to spherical outflow can be probed more thoroughly than is possible at high redshift, and any insight into the radiation mechanisms of LLGRBs can inform our understanding of the GRB population at large.

Among known LLGRB sources, the remarkably similar sources GRB 060218/SN 2006aj \citep{campana06,soderberg06,mazzali06b,kaneko06,pian06} and GRB 100316D/SN 2010bh \citep{starling11,chornock10,fan11,cano11,margutti13} stand out as unique due to their long time-scale, smooth single-peaked light curve, anomalously soft and bright X-ray afterglow, and the presence of significant thermal X-ray and optical components at early times \citep{campana06,kaneko06,starling11,margutti13}.  Several important and compelling questions concerning these two bursts remain open.  In a narrow sense, the atypical prompt emission, the origin of the X-ray blackbody component, and the unusual X-ray afterglow are hard to explain in terms of standard GRB theory.  In a broader sense, we do not know whether the progenitor system is the same as for typical GRBs: do these ultra-long LLGRBs represent the low-luminosity end of a continuum of collapsar explosions, or a different stellar endpoint altogether?  The answer to this question has important implications for high-mass stellar evolution, the connection between SNe and GRBs, and the low-energy limits of GRB physics, especially considering that events similar to GRB 060218 and GRB 100316D are likely more common than cosmological GRBs.  The peculiar nature of these bursts, the wealth of timely observations in many wavebands (especially for GRB 060218), and the lack of a consensus picture for their behaviour make these particular LLGRBs prime targets for theory.  

Accordingly, a wide variety of models have been proposed to explain the many facets of GRB 060218.  \citet{campana06} and \citet{waxman07} originally modeled the prompt X-ray emission as shock breakout from a circumstellar shell at $\sim 10^{12}$\,cm. The breakout duration in this case, assuming spherical symmetry, is only a few hundred seconds; however, \citet{waxman07} suggested asphericity as a means to lengthen the burst time-scale.  On the other hand, \citet{ghisellini07b} argued against the shock breakout interpretation, showing that fine tuning is required to bring about a large change in breakout time-scale through asymmetrical effects.  \citet{ghisellini07a} presented an alternative synchrotron self-absorption model for the prompt emission, but the high brightness temperature and small emitting area in their model are at odds with radio observations, which suggest only mildly relativistic speeds \citep{soderberg06}.  \citet{dai06} found that Compton scattering of soft input photons off relativistic external shocks driven by an inner outflow could roughly reproduce the observed prompt light curve.  In the same vein, \citet{wang07} showed that a Fermi acceleration mechanism could upscatter breakout thermal photons, creating a high energy power law tail to the thermal distribution.  However, it is unlikely that thermal equilibrium is obtained in a relativistic breakout, and photon energies are limited by Compton losses \citep{katz10,ns10,ns12}. \citet{li07} and \citet{chevalier08} investigated the prompt UV/optical emission, and demonstrated that shock breakout could reproduce the optical flux, given a large breakout radius of $5 \times 10^{13}$\,cm.  (This large radius was initially viewed as a problem; see, however, the discussion in Section \ref{SB} below.)  \citet{bjornsson08} also put forth a model for the prompt UV, based on optically thick cyclotron emission. \citet{np14} showed that an early UV/optical peak could be attributed to cooling emission from an extended low-mass circumstellar envelope shock-heated by the passage of fast SN ejecta.  They did not discuss the case of SN 2006aj, although it appears in their Figure 1 as an example of extended envelope interaction.

Another possibility for the prompt emission is that GRB 060218 is an ordinary GRB jet viewed off-axis.  However, a solely geometrical effect should result in a frequency-independent, or achromatic, break in the light curve, whereas the break in GRB 060218 is chromatic in nature \citep{amati06}.  \citet{mandeich10} considered a scenario for GRB 060218 in which primary radiation scatters off material radiatively accelerated slightly off-axis from the line of sight; this acceleration can explain the chromatic behaviour of the afterglow.  However, as their model still required an unusually soft, long-duration, and low-luminosity primary photon source, it did not give insight into the fundamental factor distinguishing LLGRBs from most bursts.

\citet{soderberg06} and \citet{fp06} tackled the X-ray and radio afterglow.  In each case, the radio could be explained by a synchrotron self-absorption in a wide ($\theta \ga 1$), mildly relativistic ($\Gamma \sim 2$--$3$) outflow, but the high X-ray afterglow flux could not be accounted for in a simple external shock synchrotron model.  \citet{soderberg06} attributed this X-ray excess to a forming magnetar, while \citet{fp06} preferred late-time fallback accretion on to a central compact object.  \citet{toma07} suggested that the radio afterglow was better explained by the late non-relativistic phase of an initially collimated jet outflow.  They inferred a jet luminosity $10^{45}$\,erg\,s$^{-1}$, an initial jet Lorentz factor $\Gamma \sim 5$, and an initial jet opening angle $\theta \sim 0.3$, and showed that a hot low-luminosity jet could successfully penetrate a WR star and expand upon breakout to achieve these initial conditions.  Based on the smooth light curve and long engine duration, they posited a neutron star-powered rather than black hole-powered central engine.   \citet{duran14} calculated the synchrotron afterglow light curves from a relativistic shock breakout, and while their model could fit the radio emission of GRB 060218, it predicted too low a flux and too shallow a temporal decay for the X-ray afterglow.  \citet{margutti14} analysed the X-ray afterglows of 12 nearby GRBs and established that GRB 060218 and GRB 100316D belong to a distinct subgroup marked by long duration, soft-photon index, and high absorption.  They proposed the possibility that these afterglows are in fact dust echoes from shells $\sim$tens of parsecs across that form at the interface between the progenitor's stellar wind and the ISM. 

Until recently, most existing models have focused on explaining a particular aspect of this burst (e.g., the prompt nonthermal emission, the radio afterglow, or the optical emission), while leaving the other components to speculation.  \citet{nakar15} recently suggested a model that attempts to unify the prompt X-rays, early optical peak, and radio emission.  In his picture, the prompt X-ray and optical emission arise from the interaction of a typical GRB jet with a low-mass envelope surrounding the progenitor star.  The short-lived jet is able to tunnel through the progenitor star, but is choked in the envelope, powering a quasi-spherical, mildly relativistic explosion akin to a low-mass SN.  The prompt X-rays are produced by the shock breaking out of the optically thick envelope \citep[as described in][]{ns12}, and optical radiation is emitted as the envelope expands and cools \citep[as in][]{np14}.  Interaction of the breakout ejecta with circumstellar material (CSM) generates the radio via synchrotron radiation \citep[as in][]{duran14}.  Nakar's model does not, however, explain the unusual X-ray afterglow or the presence of thermal X-rays at early times.

In this paper, we present a plausible alternative to Nakar's model for this peculiar burst, building on previous jet models.  In Section \ref{observations}, we give an overview of observations of GRB 060218, and discuss the key features that must be reproduced by any model.  In Section \ref{SB}, we address some problems with a straightforward shock breakout view for the prompt emission, and provide motivation for adopting a long-lived jet instead.  In Section \ref{model}, we describe how each component of the observed radiation is generated in our engine-driven model for GRB 060218, and check that our picture is self-consistent.  Advantages, drawbacks, and predictions of our model, ramifications for GRB classification, and future prospects are discussed in Section \ref{discussion}, before we conclude in Section \ref{conclusions}.

%---------------------------------------------------------------------------------------------------------------------------
% **************************************************SECTION 2**************************************************
%---------------------------------------------------------------------------------------------------------------------------

\section{Overview of observations}
\label{observations}

The X-ray evolution of GRB 060218 and GRB 100316D can be divided into a prompt phase, an exponential or steep power-law decline, and an afterglow phase.  Remarkably, these two objects share many observational features, perhaps suggesting that they have similar origins.  In both objects, we see:
\begin{itemize}
\item {Prompt nonthermal X-rays and $\gamma$-rays with a Band-like spectrum, but with lower luminosity, lower peak energy, and longer time-scale as compared to cosmological GRBs.}
\item {Thermal X-rays with roughly constant temperature $\sim 0.1$\,keV over the first $\sim1000$\,s.}  
\item {Strong thermal UV/optical emission on a time-scale of hours to days.}  
\item {A radio afterglow lasting tens of days and implying mildly relativistic outflow.}  
\item {An X-ray afterglow that is brighter and softer than expected in standard synchrotron models.}
\end{itemize} 
Any unified model for these bursts must account for each of these components.  Here we summarise multiwavelength observations during the prompt and afterglow phases of GRB 060218 and GRB 100316D.

\textbf{Prompt X-rays/$\gamma$-rays:} The nonthermal spectrum of GRB 060218 from $t=200$\,s to $t=3000$\,s is well fit by a Band function \citep{band93} with low-energy photon index $\Gamma_1 = -1.0$ and high-energy photon index $\Gamma_2 = -2.5$, implying $F_\nu \propto \nu^0$ at low energies and $F_\nu \propto \nu^{-1.5}$ at high energies.  $\Gamma_1$ and $\Gamma_2$ remain roughly constant over the evolution \citep{toma07}.  \citet{kaneko06} found a somewhat different low-energy index, $\Gamma_1 = -1.4$, when fitting the spectrum with a cut-off power law instead of a Band function.  These values are typical for long GRBs \citep{ghirlanda07}.  The peak energy $E_p$  of the best-fitting Band function decreases as $E_p \propto t^{-1.6}$ from $t=600$\,s until the end of the prompt phase.  At $700$\,s, $E_p = 10$\,keV \citep{toma07}.  Despite its low luminosity, GRB 060218 obeys the Amati correlation between $E_p$ and luminosity \citep{amati06}.  In addition to the nonthermal Band function, a significant soft thermal component was detected in the spectrum.  \citet{campana06} found that the blackbody temperature remains nearly constant at $0.17$\,keV throughout the prompt phase \citep{campana06}.  A later analysis by \citet{kaneko06} determined a slightly lower temperature, $0.14$\,keV, for times after several hundred seconds (see their Figure 7).

The prompt XRT ($0.3$--$10$\,keV) light curve of GRB 060218 can be decomposed into contributions from the thermal and nonthermal parts; the nonthermal component dominates until approximately 3000 s (\citealt{campana06}; see also Figure 1 in \citealt{ghisellini07b}).  The total (nonthermal + thermal) isotropic-equivalent luminosity in the XRT band grows as $L_{XRT} \propto t^{0.6}$ for the first $1000$\,s, when it reaches a peak luminosity $\sim 3 \times 10^{46}$\,erg\,s$^{-1}$, then declines as roughly $t^{-1}$ until $\sim 3000$\,s, fading exponentially or as a steep power law after that \citep{campana06}.  The thermal component initially comprises about $\sim 1/6$ of the total XRT band luminosity, and its light curve evolves similarly: at first it increases as a power law, rising steadily as $L_{th} \propto t^{0.66}$ \citep{liang06} until it peaks at $1 \times 10^{46}$\,erg\,s$^{-1}$ at $t = 3000$\,s \citep{campana06}.  At that time, the thermal and nonthermal luminosities are about equal, but during the steep decline phase ($3000$--$7000$\,s), the thermal component comes to dominate the luminosity, indicating that the nonthermal part must decline more steeply \citep{campana06}.  The light curve in the BAT band ($15$--$150$\,keV) is initially very similar to the XRT light curve, increasing as about $t^{0.8}$ with roughly the same luminosity.  Though its maximum luminosity ($\sim 3 \times 10^{46}$\,erg\,s$^{-1}$) is similar to the peak XRT flux, the BAT flux peaks earlier, at $t=400$\,s.  Furthermore, it decays faster after the peak, falling off as $L_{BAT} \propto t^{-2}$ from $400$--$3000$\,s \citep{campana06,toma07}.

Evidence for a blackbody spectral component has also been claimed for GRB 100316D, with a similar constant temperature $kT =  0.14$\,keV \citep{starling11}.  However, the presence of this thermal component has been called into question based on a large change in its statistical significance with the latest XRT calibration software \citep{margutti13}.  The nonthermal spectrum of this burst is similar to that of GRB 060218: its peak energy has about the same magnitude and declines in a similar fashion, and its low-energy photon index is also nearly the same over the first $\sim 1000$ seconds \citep[][see their Figure 4]{starling11}.

Compared to GRB 060218, GRB 100316D is more luminous in the XRT band, with $L_{XRT} \sim 10^{48}$\,erg\,s$^{-1}$.  In this case, the XRT light curve has nearly constant luminosity ($\propto t^{-0.13}$) for the first $800$\,s \citep{starling11}.  (For this burst, there are no X-ray data available from $800$--$30000$\,s.)  If the light curve is broken into blackbody and Band function components, the nonthermal flux strongly dominates over the thermal contribution, with $L_{XRT}/L_{th} \sim 30$.

\textbf{Optical photometry:}  From the first detection of GRB 060218 at a few hundred seconds, the UV/optical emission slowly rises to a peak, first in the UV at $31$\,ks, and then in the optical at $39$\,ks \citep{campana06}.  The light curves dipped to a minimum at around $\sim 200$\,ks, after which a second peak occurred around $800$\,ks, which can be attributed to the emergence of light from the supernova SN 2006aj.  Like other GRB-supernovae, 2006aj is a broad-lined Type Ic \citep{pian06}, but its kinetic energy $E_k \approx 2  \times 10^{51}$\,erg is an order of magnitude smaller than usual  \citep{mazzali06b}.

GRB 100316D was not detected with UVOT \citep{starling11}.  Its associated supernova, SN 2010bh, peaked at $\sim 10$\,days \citep{cano11}.  While detailed optical data is not available for the earliest times, SN 2010bh does show an excess in the B-band at $t=0.5$\,days \citep{cano11}, which is at least consistent with an early optical peak.

\textbf{X-ray/radio afterglow:}  Once the prompt emission of GRB 060218 has faded, another component becomes visible in the XRT band at $10000$\,s.  This afterglow has luminosity $L_{ag} = 3 \times 10^{43}$\,erg\,s$^{-1}$ when it first appears, and fades in proportion to $t^{-1.2}$ until at least $t=10^6$\,s.  While this power law decay is typical for GRBs, the time-averaged X-ray spectral index ($\beta$ in $F_\nu \propto \nu^{\beta}$) is unusually steep, $\beta_X = -2.2$ \citep{campana06,soderberg06}.  The measured spectral index at late times ($0.5$--$10$\,days) is $\beta_X = -4.5$ \citep{margutti14}, suggesting that the spectrum softens over time.

Radio observations of GRB 060218 beginning around $\sim 1$\,day indicate a power-law decay in the radio light curve with spectral flux $F_\nu \propto t^{-0.85}$ \citep{soderberg06}, not so different from the X-ray temporal decay and typical for GRBs.  At 5 days, the spectrum peaked at the self-absorption frequency $\nu_a \approx 3$\,GHz \citet{soderberg06}.  The radio to X-ray spectral index is unusually flat, $\beta_{RX} = -0.5$ \citep{soderberg06}.  No jet break is apparent in the radio data \citep{soderberg06}, and self-absorption arguments indicate mildly relativistic motion (see Section \ref{radio_afterglow}).

The X-ray afterglow light curve of GRB 100316D can also be described by a simple power-law decay: $L_{ag} \propto t^{-0.87}$ from $t=0.4$--$10$\,days, with X-ray luminosity $ \sim 10^{43}$\,erg\,s$^{-1}$ at $t=0.4$\,days.  Like GRB 060218, its X-ray spectrum is also very soft, with $\beta_X = -2.5$ over the period $0.5$--$10$\,days \citep{margutti14}.  Because of the gap in coverage, it is unclear precisely when the prompt phase ends and the afterglow phase begins.

GRB 100316D was detected at $5.4$\,GHz from $11$--$70$\,days, with a peak at that frequency at $t \approx 30$\,days \citep{margutti13}.  This peak comes much later than that of GRB 060218, where the $5$\,GHz peak occurred at $2$--$3$\,days \citep{soderberg06}.  The late-time radio to X-ray spectral index is $\beta_{RX} < -0.4$, comparable to GRB 060218 \citep{margutti13}.  No jet break is detected out to $66$\,days, and the estimated Lorentz factor is again mildly relativistic, $\Gamma \sim 1.5$--$2$ on day 1 \citep{margutti13}.

%---------------------------------------------------------------------------------------------------------------------------
%*********************************************SECTION 3********************************************************
%---------------------------------------------------------------------------------------------------------------------------

\section{Shock breakout or central engine?}
\label{SB}

The majority of models for the prompt X-rays of GRB 060218 fall into two categories: shock breakout \citep[e.g.,][]{campana06,waxman07,ns12,nakar15} or IC scattering of blackbody radiation by external shocks \citep[e.g.,][]{dai06,wang07}.  The latter type requires seed thermal photons for IC upscattering; while \citet{dai06} and \citet{wang07} assumed these photons were produced by a shock breakout event, other thermal sources such as a dissipative jet are also possible.  Here, we point out some difficulties with a shock breakout interpretation of the prompt X-ray emission, and suggest some reasons to consider a long-lived central engine scenario instead.

Early models for GRB 060218 \citep{campana06,waxman07} considered the case where matter and radiation are in thermal equilibrium behind the shock, and the thermal X-rays and thermal UV/optical emission arise from shell interaction and shock breakout, respectively.  However, for sufficiently fast shocks the radiation immediately downstream of the shock is out of thermal equilibrium, so the breakout temperature can be higher than when equilibrium is assumed \citep{katz10,ns10}.  In this scenario the prompt emission peaks in X-rays and the prompt spectrum is a broken power law with $F_\nu \propto \nu^0$ at low energies and $F_\nu \propto \nu^{-1.74}$ at high energies \citep{ns12}.  This is similar to the Band function spectrum observed in GRB 060218, motivating consideration of the case where the nonthermal X-rays originate from a relativistic shock breakout while the thermal UV/optical component comes from a later equilibrium phase of the breakout, as described in \citet{ns12}.  This interpretation still has possible problems.  For one, the origin of a separate thermal X-ray component is unclear in this picture.  In addition, the evolution of the prompt peak energy differs from the shock breakout interpretation.  In GRB 060218, the peak energy falls off as $t^{-1.6}$, while in the shock breakout model it declines more slowly as $t^{-(0.5-1)}$.  Consequently, while the peak energy inferred from relativistic shock breakout, $\sim 40$\,keV, is consistent with observations  at early times (less than a few hundred seconds), it overestimates $E_p$ for most of the prompt phase.  Another problem is that the optical blackbody emission is observed from the earliest time in GRB 060218, and it rises smoothly in all UVOT bands until peak.  In the nonequilibrium shock breakout scenario, thermal optical emission would not be expected until later times, when equilibrium has been attained.  A final issue with the shock breakout picture of \citet{ns12} is that it involved a stellar mass explosion.  Since only a small fraction of the energy goes into relativistic material in a standard SN explosion, the energy required for the breakout to be relativistic was extreme, $E_{SN} \ga 10^{53}$\,ergs.  This high energy is inconsistent with the unremarkable energy of the observed SN, $2 \times 10^{51}$\,ergs.  

One can also consider the case where the prompt optical emission is attributed to shock breakout, but the prompt X-rays have a different origin.   The large initial radius in this case is incompatible with a bare WR star, and initially seemed to rule out a WR progenitor \citep{li07,chevalier08}.  However, this calculation assumed that much of the stellar mass was located close to the breakout radius.  An extended optically thick region containing a relatively small amount of mass could circumvent this difficulty.  Such an envelope might be created by pre-explosion mass loss or a binary interaction.  There is mounting evidence for the existence of such dense stellar environments around other transients such as SN Type IIn \citep[][and references therein]{fransson14}, SN Type IIb \citep{np14}, SN Type Ibn \citep[e.g.,][]{matheson00,pastorello08,gorbikov14}, and SN Type Ia-CSM \citep{silverman13,fox15}.

The model of \citet{nakar15} builds on the relativistic shock breakout model of \citet{ns12}, while solving several of its problems.  \citet{nakar15} introduces a low-mass, optically thick envelope around a compact progenitor.  In his model, the explosion powering the breakout is driven not by the SN, but by a jet that tunnels out of the progenitor star and is choked in the envelope, powering a quasi-spherical explosion.  Having a large optically thick region preserves the long shock breakout time-scale, but in this case most of the mass is concentrated in a compact core.  Since the envelope mass is much smaller than the star's mass, the energy required for a relativistic breakout is reduced as compared to the model of \citet{ns12}.  This picture also provides a natural explanation for the optical blackbody component via cooling emission from the shocked envelope.  However, since the prompt X-rays still arise from a relativistic shock breakout, Nakar's model inherits some problems from that scenario as well, namely that the predicted peak energy evolution is too shallow, and the thermal X-rays lack a definitive origin.  It remains unclear, too, whether Nakar's model can account for the simultaneous observation of optical and X-ray emission at early times.

Given the possible difficulties with shock breakout, a different source for the prompt X-ray radiation should be considered.   \citet{bnp11} have shown that a central engine origin for certain LLGRBs is unlikely as their duration ($T_{90}$) is short compared to the breakout time.  However, due to their relatively long $T_{90}$, engine-driven models are not ruled out for GRB 060218 and GRB 100316D.  Furthermore, as discussed in Section \ref{observations}, the prompt X-ray/$\gamma$-ray emission of GRB 060218 shares much in common with typical GRBs.  As these similarities would be a peculiar coincidence in the shock breakout view, a collapsar jet origin for GRB 060218 is worth considering.  Motivated by this, we consider the case where the early optical emission is powered by interaction of the SN ejecta with a circumstellar envelope, but the prompt X-rays originate from a long-lived jet.

%---------------------------------------------------------------------------------------------------------------------------
%************************************************SECTION 4*************************************************
%---------------------------------------------------------------------------------------------------------------------------

\section{A comprehensive model for GRB 060218}
\label{model}

\begin{figure}
\includegraphics[width=\columnwidth]{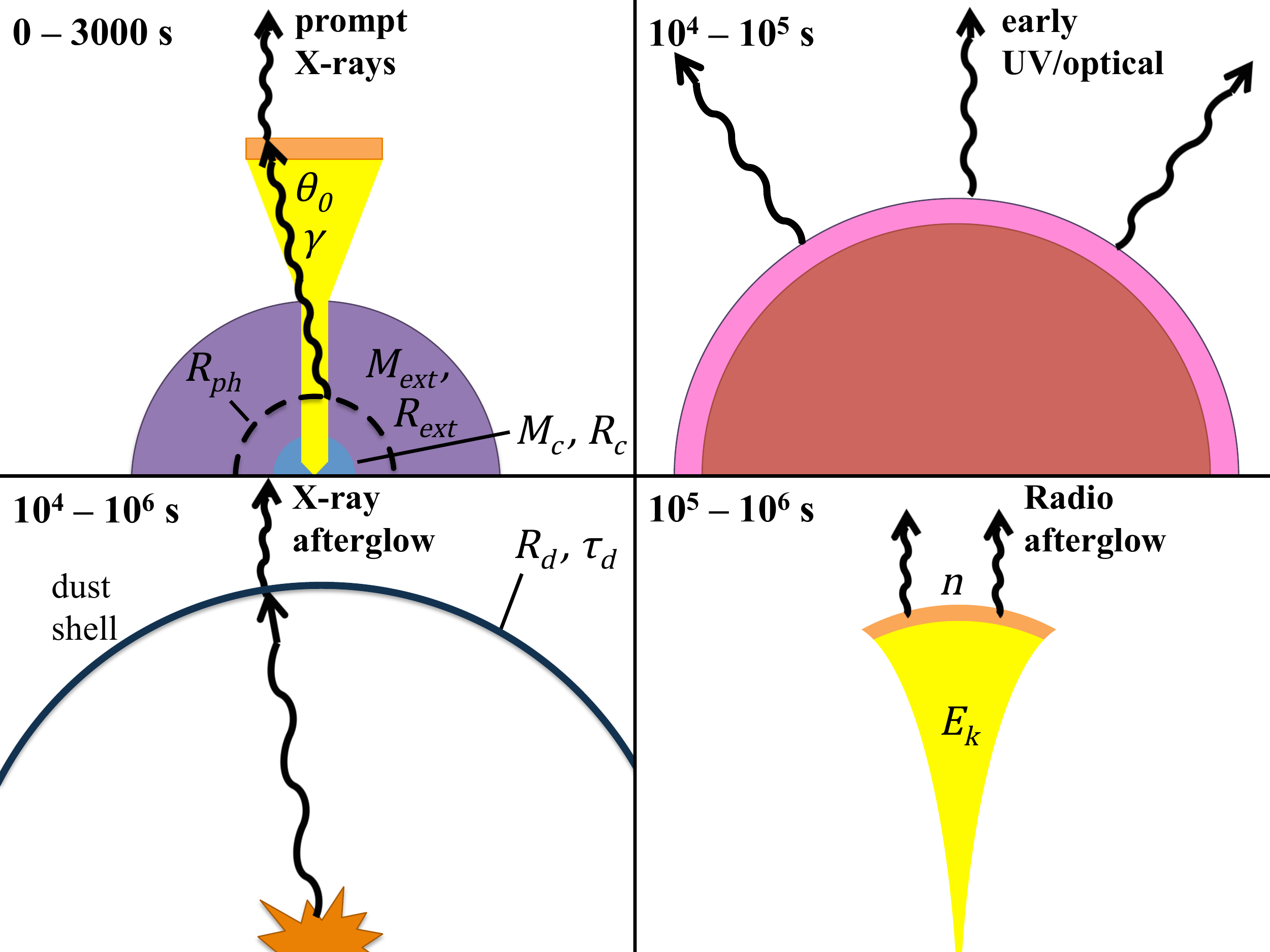}
\caption{The origin of different components of the prompt and afterglow emission in our model.  The figure is not to scale.  The progenitor has a core-envelope structure.  $M_c \sim 2\,M_\odot$ is confined to a core of $R_c \sim 10^{11}$\,cm (blue), while a mass $M_{ext} \ll M_c$ is contained mostly near the edge of an extended envelope with $R_{ext} \gg R_c$ (purple).  \textit{Upper left: } A long-lived, dissipative jet tunnels through the progenitor system.  Upon breakout, it emits blackbody radiation from radius $R_{ph}$.  Some thermal photons IC scatter from external shocks (orange) or the jet interior (yellow) to create the Band-like nonthermal component.  The jet obtains terminal opening angle $\theta_0$ and Lorentz factor $\Gamma_0$ after breakout.  \textit{Upper right: } Fast SN ejecta shock the envelope, heating it.  The slower bulk of SN ejecta (red) then lift the hot envelope (pink), which emits in the optical and UV as it expands and cools.  \textit{Lower left: } The prompt X-rays undergo scattering in a dusty region with inner radius $R_d \sim $ tens of pc and X-ray scattering optical depth $\tau_d$.  The resulting light echo outshines the synchrotron afterglow, giving rise to a characteristic soft spectrum.  \textit{Lower right: } External shock synchrotron emission from the mildly relativistic phase of the jet generates the radio afterglow.}
\label{cartoon1}
\end{figure}

A schematic of our model is presented in Figure \ref{cartoon1}.  The essential physical ingredients are a long-lived jet, an extended low-mass circumstellar envelope, and a modest amount of dust at tens of parsecs, which are responsible for the prompt X-rays/radio afterglow, early optical, and X-ray afterglow, respectively.  Below we consider the origin of each observed component in detail, and show that a reasonable match to observations can be obtained for appropriate choices of the progenitor, jet, and CSM properties.

%*****************************************prompt thermal emission********************************************
\subsection{Prompt thermal emission}
\label{prompt_thermal}

The thermal X-ray component is a puzzling aspect of GRB 060218, and it is not unique in this regard.  A recent review by \citet{peer15} lists a number of typical GRBs for which a Band $+$ blackbody model improves the spectral fit, which has been claimed as evidence for thermal emission.  \citet{burgess14} have also found evidence for thermal radiation in several other bursts.  In fact, \citet{axelsson15} have recently suggested that \textit{most} bursts must contain a broadened thermal component, because in the majority of observed bursts, the full width half maximum of the spectral peak is narrower than is physically possible for synchrotron radiation.  Although prompt thermal radiation is observationally indicated, the physical origin of this emission is yet unclear.  One possible source of thermal X-rays is a jet-blown cocoon, although the flat early light curve of GRB 060218 and GRB 100316D is hard to explain in this case \citep{peer06,starling12,ramirez02}.  Another possibility is that the blackbody emission is produced at the transparency radius in a dissipative jet outflow, as discussed in the context of GRB 060218 by \citet{ghisellini07a} and \citet{ghisellini07b}.  Here we consider the latter scenario.

\citet{pac90}, who considered photospheric emission from super-Eddington neutron star winds, showed that the photospheric radius in a spherical mildly relativistic outflow is $R_{ph} = (\dot{M}_{iso} \kappa/4 \pi \tau_{ph} \beta c) (1-\beta)$, where $\dot{M}_{iso}$ is the isotropic mass loss rate, $c$ is the speed of light, $\beta$ is the the wind velocity in units of $c$, $\kappa$ is the opacity, and $\tau_{ph} \approx 1$ is the optical depth at the photosphere.  Let $L_{th}$ and $T_0$ be the observer-frame luminosity and temperature, and define the Lorentz factor $\gamma = (1-\beta^2)^{-1/2}$.  Then the comoving luminosity and temperature are $\bar{L}_{th} = (1+\beta^2)^{-1} \gamma^{-2} L_{th}$ and $\bar{T}_0 = \gamma^{-1} T_0$.  (Here and below, a bar indicates that a quantity is measured in the frame comoving with the engine-driven outflow.)  Substituting $\bar{L}_{th}$, $\bar{T}_0$, and $R_{ph}$ into the Stefan-Boltzmann equation, one may derive $\dot{M}_{iso}$ in terms of the observables $L_{th}$ and $T_0$:
\begin{equation}
\label{Mdot}
\dot{M}_{iso} = \frac{c \tau_{ph}}{\kappa}\left(\frac{4 \pi L_{th}}{\sigma_B T_0^4} \right)^{1/2} \frac{\beta \gamma}{(1-\beta)(1+\beta^2)^{1/2}} ,
\end{equation}
where $\sigma_B$ is the Stefan-Boltzmann constant.  

Based on observations of the thermal component in GRB 060218, we take the luminosity to evolve as a power law, $L_{th} = L_0(t/t_L)^k$, before some time $t_L$, and to decline exponentially as $L_{th} = L_0 e^{(t_L-t)/t_{fold}}$ after $t_L$.  An empirical fit to the data of \citet{campana06} gives $L_0 \approx 10^{46}$\,erg\,s$^{-1}$, $t_L \approx 2800$\,s, and $t_{fold} \approx 1140$\,s.    \citet{liang06} found $k \approx 0.66$ by fitting the thermal component in the $0.3$--$2$\,keV band.   We set the temperature to a constant, $T_0$, and define $\xi = T_0/(0.17 \text{\,keV})$.  For simplicity, we assume the outflow is injected at a constant Lorentz factor; this is supported by the near constant observed temperature, as otherwise the comoving temperature would have to vary in such a way to precisely cancel the change in $\gamma$. Scaling $L_0$ to $10^{46}$\,erg\,s$^{-1}$ and $\kappa$ to $0.2$\,cm$^2$\,g$^{-1}$, and setting $\tau_{ph} =1$, the mass loss and kinetic luminosity $L_{iso} = (\gamma-1) \dot{M}_{iso} c^2$ prior to $t_L$ are
\begin{equation}
\label{Mdot2}
\dot{M}_{iso}(t) = 1.3 \times 10^{-9} \kappa_{0.2}^{-1} L_{0,46}^{1/2} \xi^{-2} (t/t_L)^{k/2} \beta \gamma^3\,M_\odot \text{\,s}^{-1}
\end{equation}
and
\begin{equation}
\label{Lj}
L_{iso}(t) = 2.3 \times 10^{45} \kappa_{0.2}^{-1} L_{0,46}^{1/2} \xi^{-2} (t/t_L)^{k/2} \beta \gamma^3 (\gamma-1) \text{\,erg\,s}^{-1}.
\end{equation}
We have assumed the Lorentz factor $\gamma$ of the outflow is large enough that the approximation $(1-\beta)^{-1}(1+\beta^2)^{-1/2} \approx \sqrt{2} \gamma^2$ applies; at worst, this differs from the exact expression by a factor $2^{1/2}$ when $\beta \rightarrow 0$.     The isotropic mass and energy of the jet are then
\begin{equation}
\label{Mj}
\begin{aligned}
M_{iso} & = \int_0^{\infty} M_{iso} dt \\
& = 3.9 \times 10^{-6} \kappa_{0.2}^{-1} L_{0,46}^{1/2} \xi^{-2} \left(\dfrac{t_{eng}}{3100 \text{ s}} \right) \beta \gamma^3 \,M_{\odot}
\end{aligned}
\end{equation}
and
\begin{equation}
\label{Ej}
\begin{aligned}
E_{iso} & = \int_0^{\infty} L_{iso} dt \\
& = 7.1 \times 10^{48} \kappa_{0.2}^{-1} L_{0,46}^{1/2} \xi^{-2} \left(\dfrac{t_{eng}}{3100 \text{ s}}\right) \beta \gamma^3 (\gamma-1) \text{\,ergs},
\end{aligned}
\end{equation}
where $t_{eng} \equiv 2t_L/(2+k) + t_{fold}$.  If the outflow is beamed into opposing jets with a small opening angle $\theta_0$, the true mass and energy of the ejected material are $M_j = (\theta_0^2/2) M_{iso}$ and $E_j = (\theta_0^2/2) E_{iso}$.  For a mildly relativistic flow ($\gamma \sim$ a few) with $\theta_0 \sim 10 \degree$, we have $M_j \sim 10^{-6}\,M_\odot$ and $E_j \sim 10^{49}$\,ergs.    The photosphere at $R_{ph} \approx 6.6 \times 10^{11} L_{0,46}^{1/2} \xi^{-2} (t/t_L)^{k/2} \gamma  \text{ cm}$ expands subrelativistically with average speed $\sim R_{ph}/t_L \sim 0.01 \gamma c$.  Note that the photosphere lies within the radius of the low-mass envelope, $R_{ext} \simeq 9 \times 10^{12}$\,cm, that we derive in Section \ref{optical} below, suggesting that dissipation occurs within the envelope. The time $t_L$ corresponds to the time when the central engine shuts off. 

In the above calculation, we have assumed for simplicity that the jet outflow is directed into an uncollimated cone.  However, as we show in Section \ref{jet_propagation}, the jet may be collimated within the envelope, and become uncollimated only after breaking out. The decollimation time-scale can be estimated as the time for the jet's cocoon to expand and become dynamically unimportant after breakout, which is $\sim R_{ext}/c_s \sim 3^{1/2}R_{ext}/c \sim 500$\,s, where $c_s$ is the sound speed.  This is short compared to the duration of prompt emission; therefore, \textit{outside} of the envelope, the assumption of an uncollimated outflow is reasonable for most of the prompt phase.  However, it appears that the photosphere is \textit{within} the envelope.  The decollimation time-scale there might be longer because it will take the jet some time to excavate the walls of the narrow hole left by its passage.  Collimation has the joint effect of decreasing the outflow's opening angle (due to the confining effect of the cocoon) and decreasing its Lorentz factor (due to more of the total jet energy going into internal versus kinetic energy).  Both of these effects lead to a smaller $M_j$ and $L_j$, for the same observed thermal luminosity and temperature.  Thus, by ignoring collimation we potentially overestimate these quantities; our derived mass loss rate and kinetic luminosity should really be viewed as upper limits.

%*******************************************extinction and absorption*****************************************
\subsection{Extinction and absorption}
\label{ext_abs}

The optical/UV extinction and the X-ray absorption to GRB 060218 are crucial for the interpretation of observations of the event, as well as giving information on gas and dust along the line of sight. The early optical/UV emission is strongly weighted to the ultraviolet, which is especially sensitive to absorption. The amount of Galactic absorption is not controversial; extinction maps of the Galaxy yield $E(B - V ) = 0.14$\,mag, while the Galactic Na I D lines indicate $E(B - V ) = 0.13$\,mag \citep{guenther06,sollerman06}.  The reddening has been estimated from the narrow Na I D lines in the host galaxy as being $E(B - V ) = 0.042$\,mag, or $A_V = 0.13 \pm 0.01$\,mag \citep{guenther06}. As noted by \citet{sollerman06}, a larger reddening is possible if there is ionization in the host galaxy. However, the properties of the host galaxy derived from fitting the spectral energy distribution and the observed Balmer line decrement point to a low extinction so \citet{sollerman06} advocate the low value obtained from the Na I D line.  Our model for the late-time X-rays (see Section \ref{xray_afterglow}) also suggests a similar low extinction.

A higher host galaxy reddening, $E(B - V ) = 0.2$\,mag, was advocated by \citet{campana06} and \citet{waxman07} because the early ($< 1$\,day) emission could be fitted by a Rayleigh-Jeans spectrum, consistent with high temperature emission. This suggestion allowed a shock breakout model for both the thermal X-ray emission and the early optical emission. This value of the reddening was also used by \citet{nakar15}, who noted that the implied blackbody temperature is $> 50,000$\,K.  \citet{nakar15} advocates the large reddening based on the slow color evolution leading up to the peak, which is expected in the Rayleigh-Jeans limit.   However, his model could in principle accommodate a smaller extinction, if the model is consistent with a constant temperature leading to the peak.

In view of the lack of direct evidence for the larger values of extinction in the host, we take the small value that is directly indicated. \citet{thone11} had derived some of the observed parameters for GRB 060218 based on Galactic extinction only.  As expected, the spectrum is then not well approximated by a Rayleigh-Jeans spectrum and a temperature in the range $30,000$--$35,000$\,K is deduced over the first half day.  A blackbody fit gives the radius at the time of peak luminosity, $10^{14}$\,cm, which yields a luminosity of $5 \times 10^{43}$\,erg\,s$^{-1}$. This can be compared to the luminosity $> 3 \times 10^{44}$\,erg\,s$^{-1}$ found by \citet{nakar15} in his larger extinction model.

The X-ray absorption column density has been obtained by fitting the observed spectrum to a model with a power-law continuum, a blackbody thermal component and interstellar absorption; \citet{kaneko06} obtain an absorbing hydrogen column density of $N_H = 6 \times 10^{21}$\,cm$^{-2}$ over 10 spectra covering the time of peak luminosity.  \citet{margutti14} infer the same absorption column from fitting an absorbed power law to the afterglow spectra.  There is no evidence for evolution of $N_H$. Using a standard conversion of $N_H$ to $A_V$ for the Galaxy, $N_H = 2 \times 10^{21} A_V$\,cm$^{-2}$ \citep[e.g.,][]{guver09}, the corresponding value of $A_V$ is 3.  There is a significant difference between the extinction determined from the Na I line and that from the X-ray absorption.

One way to reconcile the difference is to have the dust be evaporated in the X-ray absorbing region. \citet{waxman00} have discussed evaporation of dust by the radiation from a GRB; optical/UV photons with energies $1$--$7$\,eV are responsible for the evaporation. A normal burst with an optical/UV luminosity of $L_{opt} = 1 \times 10^{49}$\,erg\,s$^{-1}$ can evaporate dust out to a radius of $R_d \simeq 10$\,pc \citep{waxman00}. Since $R_d \propto L_{opt}^{1/2}$ and the peak luminosity of GRB 060218 was about $1 \times 10^{43}$\,erg\,s$^{-1}$, we have $R_d \approx 0.01$\,pc and the absorbing gas is likely to be circumstellar in origin.

%***************************************************optical*******************************************************
\subsection{UV/optical emission}
\label{optical}

Here we investigate the possibility that the optical emission is from shocked gas, but the X-ray emission is not.  We take a supernova energy of $2 \times 10^{51}$\,ergs and a core mass of $2\,M_\odot$, as determined from modeling the supernova emission (Mazzali et al. 2006b).  The optical emission has a time-scale of $\sim 1$\,day, which is characteristic of supernovae thought to show the shock breakout phenomenon \citep[see Fig. 10 in][]{modjaz09}, but the emission is brighter than that observed in more normal supernovae. As discussed in Section \ref{SB}, there is increasing evidence that massive stars can undergo dense mass loss before a supernova. We thus consider the possibility that an extended, low-mass circumstellar medium is responsible for the high luminosity.

\citet{np14} have discussed how the shock breakout process is affected by the mass of an extended envelope. When most of the stellar mass is at the radius of the surrounding envelope, a standard shock breakout, as in \citet{chevalier08}, is expected. This case applies to SN 1987A \citep{chevalier92}.  However, when the envelope mass is much less than the core mass, the early emission is determined by the emission from the envelope that is heated by the expansion of the outer part of the core. One of the distinguishing features of the non-standard case is that the red luminosity can drop with time, which is not the case for standard shock breakout. \citet{np14} note in their Fig. 1 that the early emission from GRB 060218 shows a drop in the V emission that implies the non-standard, low-mass envelope case.  Another difference is that in the standard case, the initially rising light curves turn over because the blackbody peak passes through the wavelength range of interest as the emission region cools \citep[e.g.,][]{chevalier08}, while in the non-standard case the turnover is due to all the radiative energy in the envelope being radiated and the temperature remains steady \citep{np14}.  The set of \textit{Swift}--\textit{UVOT} light curves in fact show approximately constant colors (and thus temperatures) through the luminosity peak at $\sim 3.5 \times 10^4$\,s \citep[Fig. 2 of][]{campana06}. The \textit{UVOT} observations of GRB 060218 give the best set of observations of a supernova during this early non-standard phase.

\citet{nakar15} has recently discussed the early emission from GRB 060218 in terms of interaction with a low mass envelope. The mass of the envelope was estimated at $0.01\,M_\odot$ based on the time-scale of the optical peak and an estimate of the shell velocity.  However, the expansion of the envelope was attributed to an explosion driven by the deposition of energy from an internal jet. In this case, the event is essentially a very low mass supernova. In our model, the expansion is driven by the outer, high velocity gas of the supernova explosion, as in the non-standard expansion case of \citet{np14}. The input parameters are a supernova explosion energy $E_{SN} = 2 \times 10^{51}$\,ergs and core mass $M_c = 2\,M_\odot$ \citep{mazzali06b}, a peak luminosity of $L_{p} = 5 \times 10^{43}$\,erg\,s$^{-1}$ \citep{campana06,thone11}, and a time of peak of $t_p = 3.5 \times 10^4$\,s \citep{campana06}. Since SN 2006aj was of Type Ic (no Helium or Hydrogen lines), we assumed an opacity $\kappa = 0.2$\,cm$^2$\,g$^{-1}$, appropriate for an ionized heavy element gas. These parameters can then be used to find the properties of the low mass extended envelope (subscript ext): $M_{ext} \approx 4 \times 10^{-3}\,M_\odot$, shell velocity $v_{ext} \approx 2.9 \times 10^9$\,cm\,s$^{-1}$, and energy $E_{ext} \approx 2.8 \times 10^{49}$\,ergs. These results come from the dynamics of the outer supernova layers sweeping up and out the low mass envelope around the star, and the time of the peak luminosity \citep{np14}. The value of $R_{ext} \approx 9 \times 10^{12}$\,cm is proportional to luminosity, because of adiabatic expansion.  The radius is related to the luminosity and thus the assumed absorption. These results are not sensitive to the density distribution in the extended envelope provided that most of the envelope mass is near $R_{ext}$.  The mass in the envelope derived here is sufficient that the shock wave breaks out in the envelope, as assumed in the model. At the time of maximum luminosity, the radius of the shell is $R_p \approx v_{ext}t_p = 1 \times 10^{14}$\,cm. As noted by \citet{np14}, the minimum luminosity between the two peaks of the light curve can give an upper limit to the initial radius of the core. In the case of GRB 060218, the drop in the luminosity between the peaks is shallow so that only a weak limit on the core radius can be set, $R_c \la 1.6 \times 10^{12}$\,cm.

In the choked-jet scenario, the flow must reach a quasi-spherical state prior to breaking out of the envelope semi-relativistically, which could be difficult.  Numerical simulations suggest that the bulk of the jet outflow does not become quasi-spherical until long after it becomes nonrelativistic \citep{zhang09,wygoda11,vaneerten12}.  In the lab frame, this occurs on a time-scale $\sim 5 t_{NR}$, where $t_{NR}$ is set by $\frac{4}{3} \pi \rho_{ext} (c t_{NR})^3 = E_{iso} c^{-2}$.  For a constant density envelope, we obtain
\begin{equation}
\label{t_NR}
t_{NR} \simeq 1000 \text{ s } L_{iso,51}^{1/3} M_{ext,-2}^{-1/3}
\left(\dfrac{R_{ext}}{3 \times 10^{13} \text{ cm}}\right) \left(\dfrac{t_{jet}}{20 \text{ s}}\right)^{1/3} ,
\end{equation}
where we have scaled to the values of jet isotropic luminosity, jet duration, and envelope mass and radius used by \citet{nakar15}.  As $t_{NR}$ is comparable to the time-scale $t_{ext} = R_{ext}/c \sim 1000 \left(\frac{R_{ext}}{3 \times 10^{13}\text{\,cm}}\right)$ s for a relativistic jet to break out of the envelope, the jet may only be marginally nonrelativistic at breakout.  This suggests that the breakout time $t_{br}$ is not much larger than $t_{ext}$, and it may well be that $t_{br} < 5 t_{NR}$, in which case the breakout will be aspherical.  Therefore, it is questionable whether the jet of \citet{nakar15} can become approximately spherical in the envelope, which is an assumption in his model.  As $t_{NR}/t_{ext}$ is independent of $R_{ext}$, changing the envelope's size does not help with this problem.  $t_{jet}$ can not be made much lower since it must remain larger than the time to break out of the star, which is $\sim 10$\,s in this case, and $M_{ext}$ cannot be much larger or the envelope kinetic energy would exceed the SN energy \citep{nakar15}. Thus, the problem can only be solved via a jet with lower $L_{iso}$.

These considerations show that the overall properties of the early optical/UV emission from GRB 060218 can be accounted for by a model in which there is shock breakout in a low mass, extended envelope. The model makes further predictions that can be tested in the case of GRB 060218. Approximating the observed temperature at the peak as the effective temperature leads to $T_{obs} \approx 3.5 \times 10^4$\,K, which is consistent with the observed temperature of GRB 060218 at an age of $0.085$--$0.5$\,days \citep[SI Fig. 17 of][]{thone11}. The high temperature justifies the neglect of recombination in the model.  \citet{np14} note that the optical depth of $M_{ext}$ becomes unity at $t \approx t_p(c/v_{ext})^{1/2}$, which is day 1.3 for GRB 060218; the photospheric velocity at this time gives an estimate for $v_{ext}$. The earliest spectrum of \citet{pian06} is on day 2.89, when they estimate a photospheric velocity of 26,000 km s$^{-1}$. The photospheric velocity is higher at earlier times, so there is rough agreement of the model with observations.

%*************************************************X-ray afterglow*************************************************
\subsection{X-ray afterglow}
\label{xray_afterglow}

After a steep drop, the X-ray emission from GRB 060218 enters an apparent afterglow phase at an age of $0.1$--$10$\,days. During this time, the flux spectrum is approximately a power law and the evolution is a power law in time: $F_\nu \propto \nu^{\beta_X} t^{-1.1}$ \citep{soderberg06}.  Continuous spectral softening is observed, with $\beta_X$ decreasing from $-2.2$ at $0.1$\,day to $\sim -4.5$ at $\sim 3$\,days.  The time evolution is typical of a GRB afterglow, but the spectrum is unusually steep and the indices do not obey the standard ``closure" relations for GRB afterglows \citep{fp06}. In view of this, other proposals have been made for this emission, e.g., late power from a central magnetar.  \citet{fp06} considered a wide, accretion-powered outflow as the afterglow source, but the expected light curve in that case is $F_\nu \propto t^{-5/3}$, which seems too steep to explain the observations.

In standard GRB afterglow emission, there is one population of relativistic particles that gives rise to the emission, from radio to X-ray wavelengths. However, in GRB 060218, it is difficult to join the radio spectrum with the X-rays \citep[see Fig. 1 in][]{soderberg06}; a flattening of the spectrum above radio frequencies would be necessary, as well as a sharp steepening at X-ray energies. In fact, some young supernova remnants such as RCW 86 show such spectra \citep{vink06}. The steepening would require some loss process for the high energy particles; however, \citet{soderberg06} find that synchrotron losses set in at a relatively low energy, so the observed spectrum cannot be reproduced.  In addition, the X-ray evolution does not show a jet break, as might be expected if the afterglow is produced in the external shocks of a collimated outflow.  \citet{duran14} examined a shock breakout afterglow model for the late radio and X-ray emission. They were able to model the radio emission quite adequately, but the predicted X-ray emission was considerably below that observed, decayed too slowly in time, and had the incorrect spectral index.  They concluded that the X-ray emission had some other source.

An alternative model for the emission was suggested by \citet{shao08}, that it is a dust echo of emission close to maximum light. The light curve shape expected for an X-ray echo is a plateau followed by evolution to a $t^{-2}$ time dependence. The observed light curve for GRB 060218 is between these cases, which specifies the distance of the scattering dust in front of the source, $\sim 50$ pc \citep{shao08}. \citet{shao08} applied the echo model widely to GRB light curves. However, \citet{shen09} noted two problems with this model for typical bursts. First, the required value of $A_V$ is typically $\sim$10, substantially larger than that deduced by other means. Second, the evolution is generally accompanied by a strong softening of the spectrum that is not observed.  

The case of GRB 060218 is different from the standard cases; it had a long initial burst and a large ratio of early flux to late flux. These properties are more favorable for echo emission. The early flux was $F_{pr} \approx 1\times 10^{-8}$\,erg\,cm$^{-2}$\,s$^{-1}$ lasting for $t_{pr}\sim 2000$\,s, while the late flux of $F_{late} \approx 1\times10^{-11}$\,erg\,cm$^{-2}$\,s$^{-1}$ lasted for $t_{late} \sim 20,000$\,s.  If the late emission is produced as an echo, the optical depth of the dust region is $\tau_0 = F_{late}t_{late}/F_{pr}t_{pr} = 0.01$ \citep{shen09}. The corresponding value of $A_V$ is $0.01$--$0.1$ \citep{shen09}. This value of $A_V$ is roughly consistent with that determined from the Na I D line, giving support to the echo interpretation.

To better understand the spectral softening and determine the dust properties, we numerically investigated the expected dust echo emission from a dust shell at radius $R_d$.  We used the theory of \citet{shao08}, with some modifications to specify to GRB 060218.  While \citet{shao08} assumed a flat prompt spectrum in the range $0.3$--$10$\,keV as is typical for cosmological GRBs, we instead used an empirical model including a blackbody as described in Section~\ref{prompt_thermal} and a Band function with flux and peak energy evolving according to \citet{toma07}.  In particular, the inclusion of the thermal component -- which dominates at low energies -- results in a steeper echo spectrum than predicted by \citet{shao08}.  

The parameters of the model are the dust radius $R_d$, the scattering optical depth at 1 keV $\tau_{\text{keV}}$, the minimum and maximum grain sizes $a_-$ and $a_+$, and the power-law indices $s$ and $q$ that set how the scattering optical depth per unit grain size scales with energy and grain radius, i.e. $\tau_a \propto \tau_{\text{\,keV}} E^{-s} a^{4-q}$ with $2 < s < 3$ and $3 < q < 4$ typically.  The echo flux is integrated over the range $0.3$--$10$\,keV, appropriate for the \textit{Swift} XRT band.  The parameter $a_- \approx 0.005\,\mu\text{m}$ is based on observations of galactic dust grains \citep{mathis}.  The prompt photons are approximated as being injected instantaneously at $t=0$.

\begin{figure}
\includegraphics[width=\columnwidth]{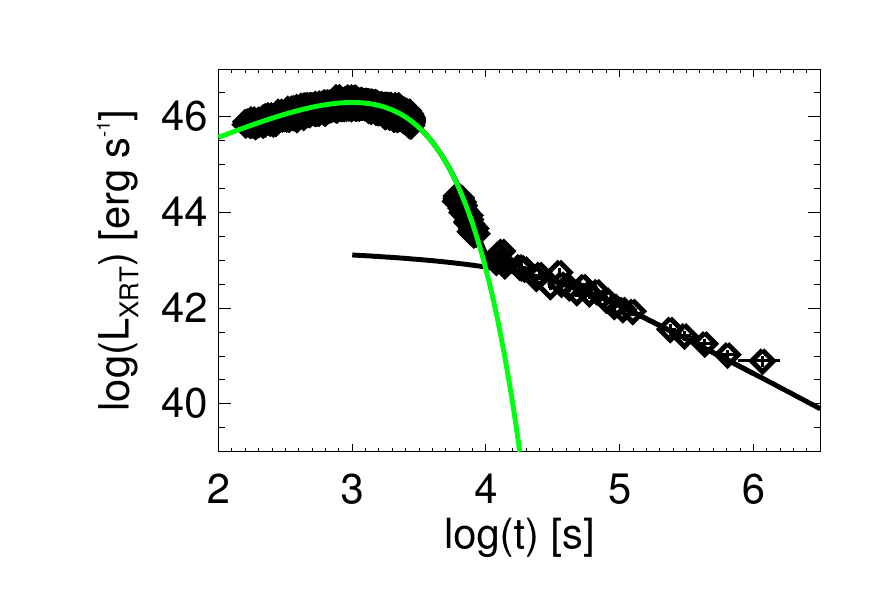}
\caption{Prompt and afterglow light curves for a dust echo model with $\tau_{\text{\,keV}} \approx 0.006$, $R_d \approx 35$\,pc, $a_+ = 0.25 \,\mu$\,m, $s=2$ and $q=4$.  The prompt data points are fit with a simple exponentially cut-off power-law, shown by the green line.  The black line indicates the contribution from dust scattering at $R_d$.}
\label{echo_LC}
\end{figure}

\begin{figure}
\includegraphics[width=\columnwidth]{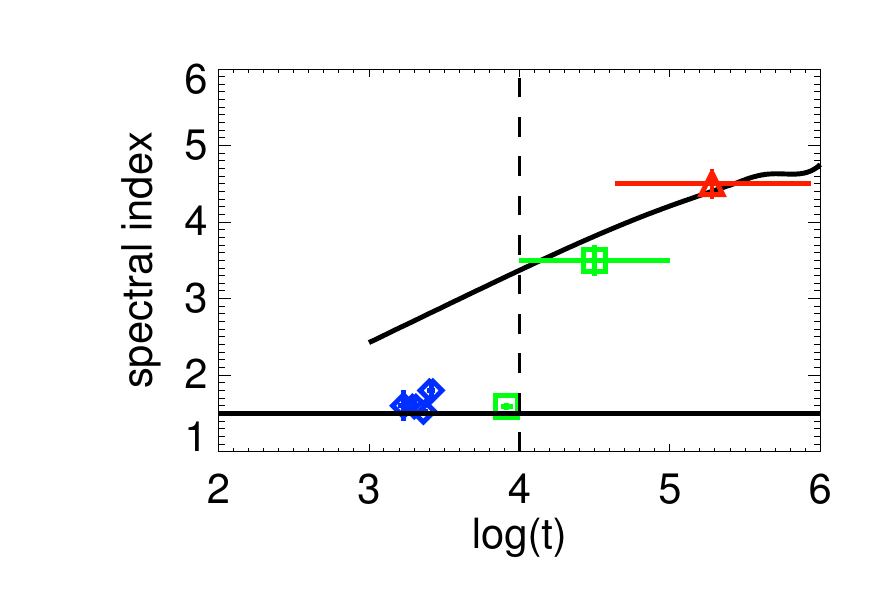}
\caption{Spectral evolution in GRB 060218.  The blue, green, and red points are taken from Table 1 in \protect\citet{toma07}, Figure 1 in \protect\citet{ghisellini07a}, and \protect\citet{margutti14} respectively.  The lower solid black line indicates the typical high-energy spectral index of the Band function, $F_\nu \propto \nu^{-1.5}$.  The upper solid black line shows the two-point XRT flux spectral index, $\log(F(10 \text{\,keV})/F(0.3 \text{\,keV}))/\log(10/0.3)$, as a function of time for our best-fitting echo model.  The time when the echo flux first exceeds the prompt flux in our model is shown by the vertical dashed line.}
\label{echo_beta}
\end{figure}

Our calculated echo light curve is shown in Fig.~\ref{echo_LC}.  We find a reasonably good fit to the light curve with reduced chi-squared of 2.1 when $\tau_{\text{\,keV}} \approx 0.006$, $R_d \approx 35$\,pc, $a_+ = 0.25\,\mu\text{m}$, $s=2$ and $q=4$.  The same model can satisfactorily reproduce the spectral evolution at late times, as depicted in Fig.~\ref{echo_beta}.    The optical depth is well-determined and robust to changes in the other parameters. There is a degeneracy between $R_d$ and $a_+$ because the afterglow flux depends only on the combination $R_d a_+^{-2}$; however, $a_+ = 0.25\,\mu\text{m}$ is roughly consistent with Galactic observations \citep{mathis,predehl}.     Varying $s$ does not greatly affect the light curve, but $s\approx2$ is preferred to match the spectral index at late times.  The model is mostly insensitive to $q$, but a larger $q$ improves the fit slightly by marginally increasing the flux at late times.  The scattering depth $\tau_{sca}$ at energy $0.8 \text{ keV} \la E \la 10 \text{ keV}$ can be converted to an optical extinction via the relation $\tau_{sca}/A_V \approx 0.15(E/1\text{\,keV})^{-1.8}$ \citep{db04}.  For $\tau_{sca} = 0.006$ at $1$ keV, $A_V \approx 0.15$, in line with Na I D line observations and the simple estimate above.   (We note that it is not necessary for these values to coincide: as the typical scattering angle is $\alpha_{sca} \sim 0.1\degree$--$1\degree$, the line of sight to the afterglow and the prompt source are separated by $\sim \alpha_{sca} R_d \sim 0.1$--$1$\,pc.  It is possible that the ISM properties could vary on this scale.)   We conclude that a moderate amount of dust located $\sim$tens of parsecs from the progenitor can plausibly explain the anomalous X-ray afterglow.

Due to the gap in observations from $1000$--$30000$\,s, the late X-ray light curve in GRB 100316D is difficult to model in detail.  None the less, some simple estimates can be made.  The prompt X-ray emission has luminosity $L_{pr} \sim 3 \times 10^{46}$\,erg\,s$^{-1}$ and time-scale $t_{pr} \sim 1000$\,s \citep{margutti13}.  The X-ray afterglow has luminosity $L_{late} \sim 2 \times 10^{43}$\,erg\,s$^{-1}$ at $t_{late} \sim 3 \times 10^4$\,s and decays as $ t^{-0.87}$ \citep{margutti13}, so $L_{late} t_{late}$ gives a reasonable estimate of the reradiated energy.  The above lead to a similar estimate for the optical depth as for GRB 060218, $\tau_d \sim 0.02$, or $A_V \sim 0.2$.  One interesting difference between the two bursts is that the spectral index of the late afterglow, $\beta_X = -2.5$, is harder in GRB 100316D than in GRB 060218 where $\beta_X = -4.5$.  (Notably, GRB 060218 is the only burst with such a steep afterglow spectrum; GRB 100316D is more typical, as other soft-afterglow bursts such as GRB 090417B and GRB 130925A also show $\beta_X \sim -2.5$ \citep{margutti14}.)  In the echo interpretation, this discrepancy can be explained partially by a difference in the prompt spectrum.  Due to the presence of a strong thermal component at low energies, the time-averaged prompt $0.3 - 10$ keV spectrum of GRB 060218 is steeper than in GRB 100316D, where the thermal component is weak and the spectrum is essentially flat at low energies.  However, this effect alone is not sufficient, as it only produces a change in spectral index of $\sim$ 1.  $R_d$ and $a_+$ also have a strong effect on $\beta_X$ because they change time-scale for spectral steepening, as does the energy dependence of the scattering cross section.  A larger $R_d$, smaller $a_+$, or lower value of $s$ (compared to our values for GRB 060218) may be necessary to obtain the correct $\beta_X$ in GRB 100316D.  However, due to a lack of data regarding the time dependence of $\beta_X$ and an insufficient light curve, we cannot say which of these effects is the relevant one.

\citet{margutti14} have recently argued that four bursts, including GRB 060218 and GRB 100316D, belong to a distinct subclass of transient taking place in a complicated CSM.  They base their claim on the unlikelihood of three unrelated properties -- high absorption column, soft afterglow spectrum, and long duration -- occurring together by chance.  They invoke a wind-swept dusty shell to account for the high X-ray absorption and steep afterglow spectrum (through an echo of the prompt emission), and propose shock breakout in a complex local CSM to explain the long duration of prompt emission, preferring this interpretation to one in which the central engine duration is intrinsically long.  Our findings support their suggestion that the very soft spectrum of GRB 060218 arises from a dust echo, but as the amount of dust in our model is not particularly high, an especially dense shell is not necessary; the dust could exist in an ISM of typical density and chemistry.  We stress that the absorption column implied by dust extinction in our model is \textit{not} consistent with the X-ray absorption column inferred from the prompt emission, as the latter is larger by a factor of $\sim 30$.  For this reason, dust scattering and X-ray absorption are unlikely to be occurring in the same place in GRB 060218.  Rather, the X-ray absorption is likely happening at small radii where dust has been evaporated.  Also, while our results do indicate a dense envelope around the progenitor star, we also differ from the \citet{margutti14} picture by adopting an intrinsically long-lived central engine.  

Our results can be compared to two other objects for which dust echo models have been proposed, GRB 130925A \citep{evans14,zhao14} and GRB 090417B \citep{holland10}.  The optical extinction inferred from modeling the afterglow as a dust echo is $A_V = 7.7$\,mag in GRB 130925A, \citep{evans14}, and in GRB 090417B it is $A_V \ga 12$\,mag \citep{holland10}.  In each case, the amount of dust required to fit the X-ray afterglow via an echo \textit{is} consistent with the absorbing hydrogen column needed to fit the X-ray spectrum \citep{evans14,holland10}.  In GRB 090417B, the high extinction can also explain the lack of an optical detection \citep{holland10}.  In contrast to GRB 060218 and GRB 100316D, GRB 130925A and GRB 090417B appear to have taken place in an unusually dusty environment, with the dust accounting for both the X-ray scattering afterglow and the large $N_H$.

Interestingly, these bursts also differ in their prompt emission.  GRB 130925A appears typical of the ultra-long class of objects described by \citet{levan13}, which also includes GRB 101225A, GRB 111209A, and GRB 121027A.  Compared to GRB 060218 and GRB 100316D, these ultra-long bursts are more luminous and longer lived, and they show variability in their light curves on short time-scales, reminiscent of typical GRBs \citep{levan13}.  The light curve of GRB 090417B is qualitatively similar to GRB 130925A, and it likewise has a longer time-scale, higher luminosity, and more variability compared to GRB 060218 \citep{holland10}.  Thus, while \citet{margutti14} have made a strong case that GRB 060218, GRB 100316D, GRB 130925A, and GRB 090417B constitute a population distinct from cosmological LGRBs, upon closer inspection GRB 130925A and GRB 090417B differ strikingly from GRB 060218 and GRB 100316D.  It seems, then, that three discrete subclasses are needed to explain their observations: 1) smooth light curve, very low-luminosity ultra-long bursts like GRB 060218/GRB 100316D, with echo-like afterglows implying a modest amount of dust; 2) spiky light curve, somewhat low-luminosity ultra-long bursts like GRB 130925A/GRB 090417B, with echo-like afterglows implying a large amount of dust; and 3) spiky-light curve bursts with typical time-scale and luminosity, and synchrotron afterglows.
 
The underlying reason why the afterglow is dominated by dust-scattered prompt emission in some cases, and synchrotron emission from external shocks in others, is unclear.  One possibility is that kinetic energy is efficiently converted to radiation during the prompt phase, resulting in a lower kinetic energy during the afterglow phase as discussed by \citet{evans14} in the context of GRB 130925A.  A second possibility is that the external shocks do not effectively couple energy to postshock electrons and/or magnetic fields.  We return to this question at the end of Section~\ref{prompt_nonthermal}.
 
%********************************************radio afterglow**********************************************
\subsection{Radio afterglow}
\label{radio_afterglow}

An essential feature of the radio afterglow in GRB 060218 is that it shows no evidence for a jet break, but instead decays as a shallow power law in time, with $F_\nu \propto t^{-0.85}$ at $22.5$\,GHz \citep{soderberg06}.  This behaviour runs contrary to analytical models of GRB radio afterglows \citep[e.g.,][]{rhoads99,sph99} which predict that, after a relatively flat decay during the Blandford--McKee phase, the on-axis light curve should break steeply to $t^{-p}$ after a critical time $t_j$.  Here, $p$ is the power law index of accelerated postshock electrons, i.e. $N(E) \propto E^{-p}$, which typically takes on values $2 < p < 3$.  The steepening is due to a combination of two effects that reduce the observed flux: when the jet decelerates to $\Gamma \sim \theta_0$, the jet edge comes into view, and also the jet begins to expand laterally.  The same general behaviour of the radio light curve is also seen in numerical simulations \citep{zhang09,vaneerten13}.  The steep decay lasts until a time $t_s$, which is the time-scale for the flow to become quasi-spherical if sideways expansion is fast, i.e. if the increase in radius during sideways expansion is negligible \citep{liviowaxman}.  While detailed simulations have demonstrated that the transition to spherical outflow is much more gradual and that the flow remains collimated and transrelativistic at $t_s$ \citep{zhang09,vaneerten12}, numerical light curves none the less confirm that analytical estimates of the radio flux that assume sphericity and nonrelativistic flow remain approximately valid for on-axis observers at $t > t_s$ \citep{vaneerten12,wygoda11}.  After $t_s$, the light curve gradually flattens as the flow tends toward the Sedov--Taylor solution, eventually becoming fully nonrelativistic on a time-scale $t_{NR}$.  Therefore, the smooth and relatively flat light curve of GRB 060218 over the period $2$--$20$\,days suggests one of two possibilities: either we observed the relativistic phase of an initially wide outflow that took $t_j \ga 20$\,days to enter the steep decay phase, or we observed the late phase of an outflow that became transrelativistic in $t_s \la 2$\,days and that may have been beamed originally.  

In either scenario, a light curve as shallow as $t^{-0.85}$ is not easily produced in the standard synchrotron afterglow model.  One issue is that such a shallow decay suggests that the circumstellar density profile and postshock electron spectrum are both flatter than usual.  Throughout the period of radio observations, the characteristic frequency $\nu_m$, the synchrotron self-absorption frequency $\nu_a$, and the cooling frequency $\nu_c$ are related by $\nu_m < \nu_a < \nu_c$ \citep{soderberg06}.  As the $22.5$\,GHz band lies between $\nu_a$ and $\nu_c$, the expected light curve slope in the relativistic case is $t^{3(1-p)/4}$ for a constant density circumstellar medium, and $t^{(1-3p)/4}$ for a wind-like medium \citep{lev12,fp06}.  In the nonrelativistic limit, the slopes are $t^{3(7-5p)/10}$ (constant density) and $t^{(5-7p)/6}$ (wind) \citep{lev12}.  In order to fit the observed slope $t^{-0.85}$, we require a constant density medium and $p = 2.1$ (relativistic) or $p=2.0$ (nonrelativistic).  However, \citet{pk02} found that the afterglows of several typical GRBs were best explained with a constant density model, and a low $p$-value was indicated for a number of bursts in their sample.  Hence, GRB 060218 does not seem so unusual in this regard.

A second point of tension with the shallow light curve is the observed Lorentz factor.  \citet{soderberg06} inferred a mildly relativistic bulk Lorentz factor $\Gamma \simeq 2.4$ from an equipartition analysis.  However, they based their analysis on the treatment of \citet{kulkarni98}, which did not include the effects of relativistic expansion.  A more accurate calculation that takes relativistic and geometrical effects into account was carried out by \citet{duran13}.  From Figure 2 in \citet{soderberg06}, we estimate that, at day 5, the spectral flux at peak was $F_p \sim 0.3$\,mJy and the peak frequency was $\nu_p = \nu_a \sim 3$\,GHz.  Applying equation (5) in \citet{duran13}, we obtain a bulk Lorentz factor $\Gamma \approx 0.8$.   On the other hand, using their equation (19) for the equipartition radius $R_{N}$ in the nonrelativistic limit, we find $\beta \sim R_N/ct \approx 1.3$.  These results indicate that the outflow is in the mildly relativistic ($\beta \Gamma \sim 1$) limit, where neither the Blandford--McKee solution (which applies when $\Gamma \gg 2$) nor the Sedov--Taylor solution (which applies when $\beta \ll 1$) is strictly valid.  As discussed above, one expects a relatively shallow light curve slope in these limits, but during the transrelativistic transitional regime the slope tends to be steeper.  In spite of these caveats, we press on and compare the relativistic and nonrelativistic limits of the standard synchrotron model.

The possibility of a wide, relativistic outflow was first considered by \citet{soderberg06}.  Their spherical relativistic blast wave model predicts an ejecta kinetic energy $E_{k} \sim 2 \times 10^{48}$\,ergs and a circumburst density $n \sim 100$\,cm$^{-3}$, assuming fractions $\epsilon_e \sim 0.1$ and $\epsilon_B \sim 0.1$ of the postshock energy going into relativistic electrons and magnetic fields, respectively.   In order to postpone the jet break, they presumed the initial outflow to be wide, with $\theta_0 \ga 1.4$ \citep{soderberg06}.  Yet, as \citet{toma07} pointed out, given the isotropic equivalent $\gamma$-ray energy $6 \times 10^{49}$\,ergs, the parameter set of \citet{soderberg06} predicts an unreasonably high $\gamma$-ray efficiency, $\eta_\gamma \approx 98\%$.  \citet{fp06} refined this analysis and showed that parameters $E_{k} \sim 10^{50}$\,ergs, $n \sim 100$\,cm$^{-3}$, $\epsilon_e \sim 10^{-2}$, and $\epsilon_B \sim 10^{-3}$ also fit the data while keeping the $\gamma$-ray efficiency within reason, but the origin of the spherical (or very wide) relativistic outflow is still unclear.  One possibility is that the some fraction of the SN ejecta is accelerated to relativistic speeds.  However, \citet{tmm01} have found that, even for a large SN energy $\sim 10^{52}$\,ergs, only a fraction $\sim 10^{-4}$ goes into relativistic ejecta.   It therefore seems implausible that $\sim 5\%$ of the SN energy $2 \times 10^{51} \text{ ergs}$  could be coupled to relativistic material in GRB 060218.  A choked jet in a low-mass envelope, as discussed by \citet{nakar15}, provides an alternative way to put significant energy into a quasi-spherical, relativistic flow.

Given the difficulties with the relativistic scenario, \citet{toma07} considered the possibility that the radio emission comes from the late spherical phase of an originally collimated outflow instead.  With the same assumption of $\epsilon_e = \epsilon_B = 0.1$, \citet{toma07} infer the same kinetic energy and circumstellar density as \citet{soderberg06}.  The advantage of their view is that it eliminates the efficiency problem, as the isotropic equivalent kinetic energy during the early beamed phase is larger by a factor $2/\theta_0^2$.  

\citet{duran14} also looked at a mildly relativistic synchrotron model in the context of SN shock breakout.  In this case, the light curve decays more slowly since energy is continuously injected as the outer layers of the SN ejecta catch up to the shocked region.  As a result, the radio light curve is better fit by a wind profile than a constant density in the breakout case \citep{duran14}.  Their study adopts a fixed $\epsilon_e= 0.2$, and a fixed energy and Lorentz factor for the fast shell dominating breakout emission, $E_f = 2 \times 10^{50}$\,ergs and $\gamma_f = 1.3$, which are derived from the relativistic breakout model of \citet{ns12}.  They then vary $\epsilon_B$ and the wind density parameter $A_*$, concluding that $\epsilon_B = 1.5 \times 10^{-4}$ and $A_*=10$ give the best fit.  Due to degeneracy, however, other parameter sets with different energy and $\epsilon_e$ may fit the radio light curves as well. 

Unfortunately, such degeneracies involving the unknown quantities $\epsilon_e$ and $\epsilon_B$ are an unavoidable limitation when deriving $E_k$ and $n$ in the standard synchrotron model.  The available observations give only the specific flux $F_\nu$, the self-absorption frequency $\nu_a$, and an upper limit on the cooling frequency $\nu_c$, which is not sufficient to uniquely determine the four model parameters.  In practice, this is typically addressed by fixing two of the parameters to obtain the other two.  (For example, \citet{soderberg06} choose $\epsilon_e$ and $\epsilon_B$; \citet{duran14} fix $\epsilon_e$ and $E_k$.)  We take a different approach.  In this section and Section \ref{jet_propagation}, we consider a number of constraints from dynamics, time-scales, and direct radio, optical, and X-ray observations, assuming that the emission is from the late phase of an initially collimated jet.  We apply these conditions to constrain the available $(E_k, n , \epsilon_e, \epsilon_B, \theta_0, \gamma)$ parameter space.  We then consider whether any reasonable parameter set is consistent with a jet that could produce the observed thermal X-rays through dissipation at early times, as described in Section \ref{prompt_thermal}.
 
We begin with the constraints inferred directly from radio observations.  We have $\nu_a \sim 4 \times 10^{9}$\,Hz at 5 days, $F_\nu(22.5 \text{\,GHz}) \sim 0.25$\,mJy at 3 days, and $\nu_c \la 5 \times 10^{15}$\,Hz so that the synchrotron flux remains below the observed X-ray afterglow flux throughout observations \citep{soderberg06}.  Lower limits on $E_k$ and $n$ can be deduced by assuming $\epsilon_e  < 1/3$ and $\epsilon_B < 1/3$.  For a relativistic blast wave with $p=2.1$, we have $\epsilon_{B,-2}^{0.34} \epsilon_{e,-1}^{0.36} C_p^{0.36} E_{k,51}^{0.34} n^{0.33} \sim 0.44$, $\epsilon_{B,-2}^{0.78} \epsilon_{e,-1}^{1.1} C_p^{1.1} E_{k,51}^{1.28} n^{0.5} \sim 0.0032$, and $\epsilon_{B,-2}^{-1.5}  E_{k,51}^{-0.5} n^{-1} \la 0.43$ \citep{fp06}, where $C_p = 13(p-2)/3(p-1) \approx 0.39$.  In this case, we find $E_k \ga 7 \times 10^{47}$\,ergs and $n \ga 3$\,cm$^{-3}$.  Similarly, in the nonrelativistic limit \citet{toma07} derived $\epsilon_{B,-2}^{1/3} \epsilon_{e,-1}^{1/3} E_{k,51}^{1/3} n^{1/3} \sim 1$, $\epsilon_{B,-2}^{3/4} \epsilon_{e,-1} E_{k,51}^{1.3} n^{0.45} \sim 0.003$, and $\epsilon_{B,-2}^{3/2} E_{k,51}^{3/5} n^{0.9} \ga 0.4$.  This leads to the constraints $E_k \ga 1 \times 10^{47}$\,ergs and $n \ga 60$\,cm$^{-3}$.
 
The minimum synchrotron energy $E_{min}$ provides a further constraint on burst energetics.  In general, calculating $E_{min}$ requires integrating the specific synchrotron luminosity $L_\nu$ over a range of frequencies $\nu_{min}$--$\nu_{max}$.  However, when $p \simeq 2.5$, the dependence of $E_{min}$ on $\nu_{min}$, $\nu_{max}$, and $p$ is weak \citep{longair}.  In that case, if $L_\nu$ is measured at frequency $\nu$, one can obtain a rough estimate of $E_{min}$ by setting $\nu = \nu_{min}$: with quantities given in cgs units, $E_{min} \simeq 8.0 \times 10^{13} (1+\eta)^{4/7} V^{3/7} \nu^{2/7} L_\nu^{4/7}$\,ergs \citep{longair}, where $\eta$ is the ratio of proton energy to electron energy, which is not known.  \citet{soderberg06} estimated that the size of the radio-emitting region is $R = 3 \times 10^{16}$\,cm at $t=5$ days, so the emitting volume at that time can be approximated by $V \sim \frac{4}{3} \pi R^3 \sim 1.1 \times 10^{50}$\,cm$^3$.  At the same time, the flux density at $\nu = 4.86$\,GHz was $S_\nu = 300 \,\mu\text{Jy}$, implying a specific luminosity $L_\nu = 7.5 \times 10^{27}$\,erg\,s$^{-1}$\,Hz$^{-1}$ given the distance $D=145$\,Mpc.  With these parameters, we find $E_{min} \sim 1.2 \times 10^{47} (1+\eta)^{4/7}$\,ergs.  Compared to the above estimate, this puts a stricter lower limit on the energy when $\eta$ is large.

A further condition comes from time-scale considerations, since the steep $t^{-p}$ part of the light curve should fall outside of the observational period.  For an on-axis observer, a numerically calibrated expression for the jet break time in the observer frame is $t_j = 3.5 E_{iso,53}^{1/3} n^{-1/3} (\theta_0/0.2)^{8/3}$\,days \citep{vaneerten10b}.  In the relativistic case, we need $t_j \ga 20$\,days, so $E_{k,51}^{1/3} n^{-1/3} (\theta_0/0.2)^2 \ga 5.7$.  On the other hand, the time $t_s$ that roughly marks the end of the steep light curve phase is $t_s \simeq 365 E_{iso,53}^{1/3} n^{-1/3} (\theta_0/0.2)^{2/3}$\,days \citep{liviowaxman}.  Since $t_s \la 2$ is needed for the nonrelativistic model, we have $E_{k,51}^{1/3} n^{-1/3} \la 0.0055$.  Note that $t_j \sim (\theta_0^2/4)t_s$.

For typical burst energies and CSM densities, the relativistic scenario requires a very wide opening angle to make $t_j$ sufficiently large.  For example, the parameters of \citet{soderberg06} require $\theta_0 \ga 80 \degree$.  An equally large $\theta_0$ is inferred for GRB 100316D.  In that object, the radio afterglow has a similar slow temporal decay, but the time-scale of the $F_\nu$ peak at $8.5$\,GHz was $\sim 10$ times longer, occurring at $30$\,days \citep{margutti13} as compared to $3$\,days in GRB 060218 \citep{soderberg06}, and the radio luminosity is about 10 times higher at $20$\,days \citep{margutti13}.  Assuming the same microphysics, this implies about the same burst energy, but a circumstellar density that is higher by a factor of $100$--$1000$ \citep{margutti13}, even for a quasi-spherical outflow.  It seems unusual that the progenitors of these similar bursts have such different circumstellar environments.  In addition, the higher density leads to a \textit{smaller} $t_j$ than in GRB 060218, while radio observations show a flat light curve over the period $20$--$70$\,days \citep{margutti13} implying $t_j \ga 70$\,days, \textit{larger} than GRB 060218.  This problem is alleviated by considering a wind-like medium as in \citet{margutti13}, but in that case the expected light curve is $\propto t^{-3/2}$ for $p=2$, which seems too steep to fit observations unless one adopts $p < 2$.  

One can consider the nonrelativistic case instead, but due to the weak dependence on $E_k$ and $n$, the condition on $t_s$ is also hard to satisfy unless the burst energy is extremely low or the CSM is extremely dense.  In addition, because the flow is still highly aspherical at $t \sim t_s$, the model light curve slope will be too steep if $t \ga t_s$ only holds marginally, even if the flux is approximately correct.  The slope does not settle to the limiting Sedov--Taylor value until the outflow sphericizes, which according to numerical simulations does not occur until $\sim 5 t_{NR} \simeq 4700 E_{k,51}^{1/3} n^{-1/3} (\theta_0/0.2)^{-2/3} \text{\,days} \gg t_s$ \citep{zhang09,vaneerten12}.  

However, so far we have assumed that the CSM near the progenitor star is the same as the CSM at $\sim$ a few $10^{16}$\,cm where the radio is emitted.  It is possible that the circumstellar environment is more complicated, and in particular that the CSM density is higher closer to the progenitor star.  In fact, there is some evidence that this is the case.  The X-ray absorption column, $N_H \approx 6 \times 10^{21}$\,cm$^{-2}$, measured during the prompt phase is higher than one would expect for a constant density medium with $n \sim 100$\,cm$^{-3}$, even if that medium extended to $\sim 10$\,pc scales.  Thus, we speculate that the shell emitting the radio could have undergone additional deceleration by sweeping up the material responsible for X-ray absorption at some time $t < 2$\,days.  While the absorbing column through the expanding outflow is expected to change with time, the measured $N_H$ is constant during the prompt phase.  Therefore, the bulk of the absorbing material would have to lie outside the jet throughout the first $10^4$\,s.  Note that $N_H$ also stays constant throughout the X-ray afterglow  from $10^4$--$10^6$\,s, but this does not provide any additional constraints on our model, since the afterglow in our picture is a light echo and thus inherits the absorption of the prompt component (see Section \ref{xray_afterglow}).   Hence most of the absorbing mass must be confined to the radial range $R(10^4 \text{\,s}) < R_{abs} < R(2 \text{\,days})$.  Taking $R \sim 2 \Gamma^2 c t$, where $\Gamma$ is the Lorentz factor of the forward shock, we find $0.002 \text{\,pc} \la R_{abs} (\Gamma/3)^{-2} \la 0.03 \text{\,pc}$.  The total mass of absorbing material, assuming it is distributed isotropically, is $M_{abs} \sim 4\pi R_{abs}^2 N_H m_p$, so $0.002\,M_\odot \la M_{abs} (\Gamma/3)^{-4} \la 0.5\,M_\odot$.  Note that less absorbing mass is necessary if it is located closer to the star, or if it is distributed preferentially along the poles.    The jet will sweep a mass $M_{sw} \sim (\theta_0^2/2) M_{abs}$, which is sufficient to decelerate it to nonrelativistic speeds if $M_{sw} \gg M_j$.  From equation (\ref{Mj}), $M_j \sim 10^{-4} (\theta_0^2/2) (\gamma/3)^3 M_\odot$ for the parameters of GRB 060218, thus we find
\begin{equation}
7 \Gamma (\Gamma/\gamma)^3 \la M_{sw}/M_j \la 2000 \Gamma (\Gamma/\gamma)^3.
\end{equation}
$M_{sw}/M_j > 1$ is possible to satisfy for $\Gamma \ga  0.1 \gamma^{4/3}$, and is always satisfied for $\Gamma \ga 0.6 \gamma^{4/3}$.  Therefore, for the mildly relativistic case we consider, it seems plausible that the mass responsible for the X-ray absorption could also be responsible for decelerating the jet.  

\begin{figure}
\centering
\begin{tabular}{cc}
\includegraphics[width=0.5\textwidth]{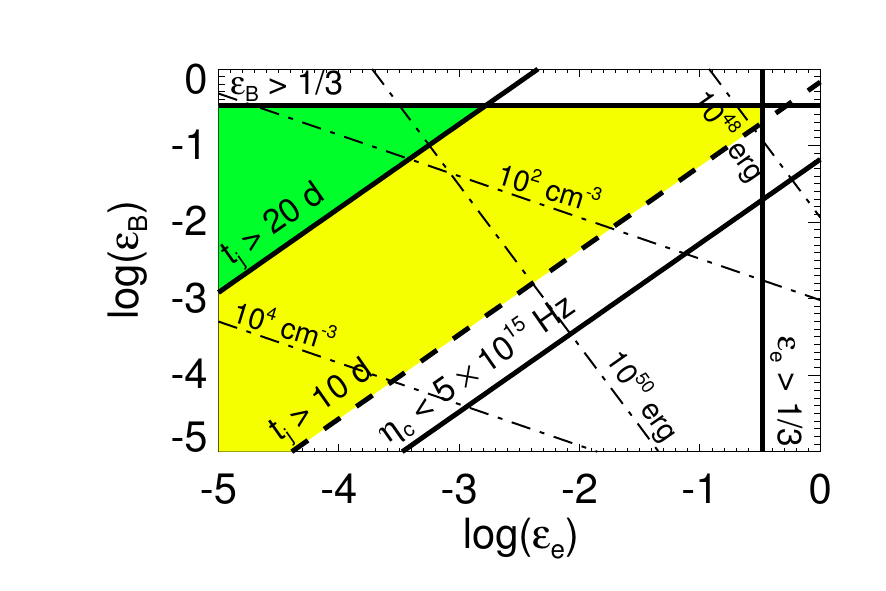} \\
\includegraphics[width=0.5\textwidth]{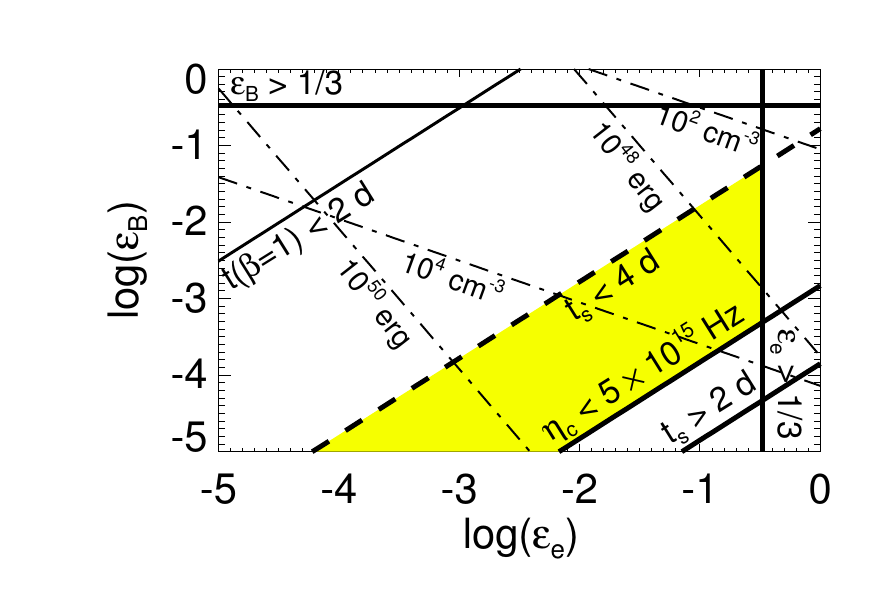} \\
\includegraphics[width=0.5\textwidth]{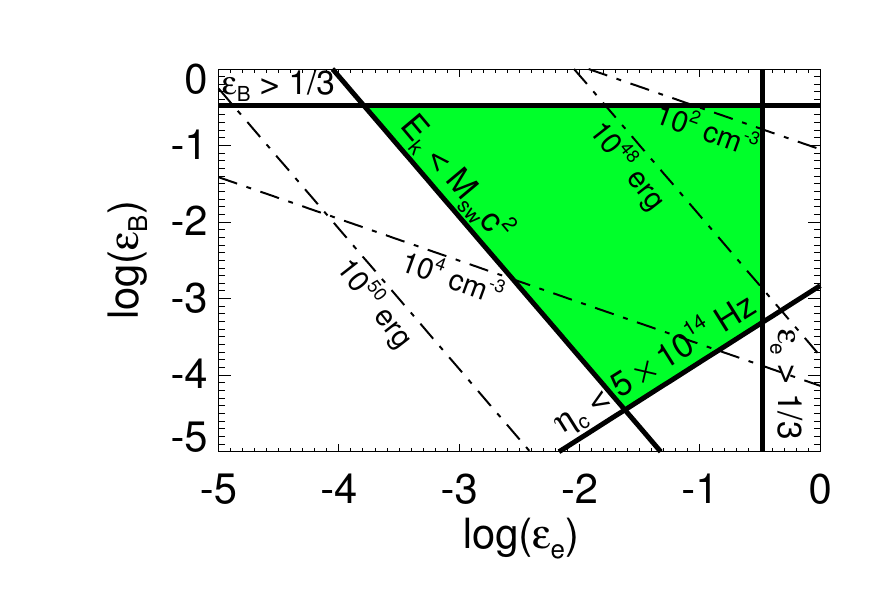}
\end{tabular}
\caption{Constraints described in Section \ref{radio_afterglow}, depicted in the $\epsilon_e$--$\epsilon_B$ plane.  In each plot, thin dash-dotted lines of constant $E_k$ ($10^{48}$\,ergs and $10^{50}$\,ergs) and $n$ ($10^2$\,cm$^{-3}$ and $10^4$\,cm$^{-3}$) are drawn.  The conditions $\epsilon_B = 1/3$, $\epsilon_e = 1/3$, and $\nu_c = 5 \times 10^{15}$\,Hz are shown as heavy solid lines, as labelled in the diagram.  Regions where all conditions are met are shaded in green, while yellow regions indicate that the conditions are met if the time-scale constraints are relaxed by a factor of 2.  \textit{Top:} The relativistic case with $p=2.1$.  The condition $t_j = 20$\,days is shown, assuming a wide jet ($\theta_0 \simeq 1.4$). \textit{Middle:} The nonrelativistic case with $p=2.0$, assuming the same density for all $r > R_{ext}$, with the conditions $t_s = 2$\,days, $t_s = 4$\,days, and $t(\beta \simeq 1) = 2$\,days.  Note that $t_s \la 2$\,days and $\nu_c <5 \times 10^{15}$\,Hz cannot be jointly satisfied.  \textit{Bottom:} The nonrelativistic case, assuming some additional mass $M_{sw} = 10^{-5}\,M_\odot$ is swept up prior to $2$\,days. See the text for discussion.}
\label{radioconstraints}
\end{figure}

The conditions $\epsilon_e < 1/3$, $\epsilon_B < 1/3$, $E_k > E_{min}$, $\nu_c < 5 \times 10^{15}$ Hz, and the constraint on $t_j$ (or $t_s$) can be conveniently expressed in the $\epsilon_e$--$\epsilon_B$ plane.  The result is shown in Figure \ref{radioconstraints}.    (We do not plot the line $E_k = E_{min}$ as it is largely irrelevant as long as $\eta \sim 1$.)   The first two panels show the standard, constant CSM density case, for a relativistic flow with $p=2.1$ and a nonrelativistic flow with $p=2.0$, respectively.  In the relativistic case, if the jet is very wide ($\theta_0 \simeq 1.4$), the $t_j$-condition can be marginally satisfied ($t_j \ga 10$\,days) for $\epsilon_B \ga \epsilon_e$ as depicted in the top panel.  However, because $t_j$ is sensitive to $\theta_0$, the available parameter space rapidly shrinks when $\theta_0$ is reduced: the green region disappears from the plot when $\theta_0 \la 1.0$, and the yellow region when $\theta_0 \la 0.7$.  Thus, the relativistic scenario disfavors a tightly collimated outflow for any sensible combination of $\epsilon_e$ and $\epsilon_B$.   In the nonrelativistic case, shown in the middle panel, we see that the constraints $t_s < 2$\,days and $\nu_c < 5 \times 10^{15}$\,Hz can not be jointly satisfied for any choice of parameters.  At best, the $t_s$-condition can be met marginally ($t_s \la 4$\,days) if $10^{-3} \la \epsilon_B/\epsilon_e \la 10^{-1}$.  (Here, we also show the condition $t(\beta=1) < 2$\,days, which was used by \citet{toma07}.  However, as discussed above and in \citet{wygoda11}, the radio flux still deviates considerably from the Sedov--Taylor prediction at this time because the outflow is semirelativistic.)  The situation changes if some additional mass is swept up by the jet before the first radio observation, in which case the condition $E_k < M_{sw} c^2$ replaces the upper limit on $t_s$.  This scenario is shown in the bottom panel, assuming $M_{sw} = 10^{-5}\,M_\odot$ (corresponding to an isotropic mass $10^{-3}\,M_\odot$ for $\theta_0=0.2$).  Compared to the standard cases, this case accommodates a larger set of possible parameters.  The effect of increasing (decreasing) $M_{sw}$ is to increase (decrease) the size of the green region by moving the critical line $E_k = M_{sw} c^2$ towards the lower left (upper right).

%*************************************************jet propagation***************************************************
\subsection{Jet propagation}
\label{jet_propagation}

 We now examine the evolution of the jet as it drills the star and breaks out into the surrounding medium.  For our picture so far to be plausible, several conditions must be met.  First, the initial kinetic energy of the outflow $E_{iso}$ must exceed the prompt isotropic radiated energy $E_{\gamma,iso} = 6 \times 10^{49}$\,ergs, i.e. the radiative efficiency $\eta_\gamma = E_{\gamma,iso}/E_{iso} < 1$.  Using Equation \ref{Ej}, this implies $\gamma > 2.1$.  Note that $E_{iso} = E_{\gamma_iso} + E_{k,iso}$, where $E_{k,iso} = (2/\theta_0^2) E_k$.  Second, the total breakout time from the stellar core and extended envelope, $t_b=t_{b,*} + t_{b,ext}$, should be shorter than the duration of prompt X-rays $t_L$.  Third, the interaction with the extended envelope should be dominated by the supernova, and not by the jet or cocoon.  In other words, the jet/cocoon system should not sweep up or destroy the envelope before the supernova has a chance to interact with it.  Finally, we expect that the energy in relativistic ejecta will be less than the SN energy: $E_j < E_{SN}$.

In what follows, we scale the collimation-corrected jet luminosity $L_j$ to $10^{46}$\,erg\,s$^{-1}$, corresponding to a jet energy $E_j \sim L_j t_L \sim 3 \times 10^{49}$\,ergs.  We assume a constant jet luminosity for simplicity.  (A time-varying luminosity does not affect our general conclusions, as long as the average value of $L_j$ remains the same.)  

In order for such a low-luminosity jet to penetrate the progenitor star, \citet{toma07} found that it must be hot and have a narrow opening angle $\theta_j \la 0.03$, conditions that are satisfied by a collimated jet.  The general theory of jet propagation in the collimated and uncollimated regimes was put forth by \citet{bromberg}.  Their model is applicable when the jet is injected with a Lorentz factor $\gamma_{inj}$ and opening angle $\theta_{inj}$ that satisfy $\gamma_{inj} \ga \theta_{inj}^{-1}$.  They showed that the jet is collimated if $L_j < \pi r_h^2 \rho_a c^3 \theta_{inj}^{2/3}$, where $r_h$ is the radius of jet's head and $\rho_a$ is the density of the ambient medium.  For a typical WR star with mass $M_* = 10 M_\odot$ and radius $R_* = 10^{11}$ cm, the jet is collimated for $L_j \la 10^{52} \text{ erg s}^{-1}$ \citep{bromberg}, so we are well within this regime.  While propagating in the star, collimation by the uniform-pressure cocoon keeps the jet cross section approximately constant, and the Lorentz factor below the jet head is $\gamma_j \sim \theta_{inj}^{-1}$, independent of the injection Lorentz factor $\gamma_{inj}$ \citep{bromberg}.  Later, once the jet breaks out into a low-density medium, it becomes uncollimated and its opening angle and Lorentz factor tend towards $\sim \theta_{inj}$ and $\sim \gamma_{inj}$, respectively \citep{bromberg}.  Therefore, the values $\theta_0$ and $\gamma$, which describe the jet post-breakout, provide an estimate of the injection conditions at much smaller radii, i.e. $\theta_{inj} \sim \theta_0 $ and $\gamma_{inj} \sim \gamma$. As a result,
\begin{equation}
\label{gamma_lower}
\gamma \ga \theta_0^{-1}
\end{equation}
will hold after adiabatic expansion.

In the strongly collimated limit the jet head moves nonrelativistically with speed $\beta_h \simeq L_j^{1/5} \rho_a^{-1/5} t^{-2/5} \theta_{inj}^{-4/5} c^{-1}$ \citep{bromberg}.  Let the stellar density profile be $\rho_a = \rho_0 (r/R_*)^{-\delta_*}$, with $\rho_0 = \frac{(3-\delta_*)M_*}{4\pi R_*^3}$. Typically, $1.5 \la \delta_* \la 3$ for WR stars \citep[e.g.,][]{mm99}.   Equation (B-2) in \citet{bromberg} gives the radius of the jet head as a function of time for the case of a nonrelativistic head; substituting $r=R_*$ and $\rho_a = \rho_0$ into that expression, we find the breakout time
\begin{equation}
\label{tb_star}
t_{b,*} = 94 \text{ }  C(\delta_*) L_{j,46}^{-1/3} M_{c,1}^{1/3} R_{c,11}^{2/3} \left(\dfrac{\theta_{0}}{0.2}\right)^{4/3} \text{\,s}.
\end{equation}
The order-unity constant $C(\delta_*) = \left[\frac{3}{7}\frac{5-\delta_*}{3-\delta_*}\right]^{-2/3}$ scales the result to $\delta_* = 1.5$, as this gives the most conservative estimate of the breakout time for the typical range of $\delta_*$.

Breakout from the low-mass envelope proceeds similarly to breakout from the stellar core, the main differences being that the jet head is faster and harder to collimate due to the lower ambient density.  For an envelope density profile $\rho_{ext} \propto r^{-\delta_{ext}}$, $\rho_{ext}(R_{ext}) = \frac{(3-\delta_{ext}) M_{ext}}{4 \pi R_{ext}^3}$.  While $\delta_{ext}$ is not known in general, requiring that the density decreases outwards and that most of the envelope mass is at large radii restricts its value to $0 \leq \delta_{ext} \leq 3$.  The collimation condition at the edge of the envelope can be rewritten as
\begin{equation}
\label{envelope_collimation}
L_{j,46} \la 46 (3-\delta_{ext}) M_{ext,-3} R_{ext,13}^{-1} \left(\dfrac{\theta_0}{0.2}\right)^{2/3} \equiv L_{coll,46}.
\end{equation}
For our parameters, we find that the jet remains collimated throughout the envelope, though we note that the high-luminosity jet of a typical GRB would be uncollimated in the same envelope \citep[as discussed by][]{nakar15}.  The parameter $\tilde{L} = (L_j/\rho_{ext} t^2 \theta_0^4 c^5)^{2/5}$ then determines if the jet head is relativistic ($\tilde{L} > 1$) or Newtonian ($\tilde{L} < 1$) \citep{bromberg}.  The condition for a nonrelativistic head at breakout is
\begin{equation}
\label{envelope_nonrelativistic}
L_{j,46} \la 6.9 \times 10^{-2} (3-\delta_{ext}) M_{ext,-3} R_{ext,13}^{-1} \left(\dfrac{\theta_0}{0.2}\right)^{4}.
\end{equation} 
For the range of parameters we consider, the jet is usually relativistic at the time of breakout; however, for a low luminosity or a somewhat wide opening angle, it is possible that the jet head breaks out nonrelativistically.

Since we are interested in a lower bound on the jet luminosity, we compute the breakout time in the nonrelativistic limit.  In this case, we can reuse equation (\ref{tb_star}) with $R_*$ and $M_*$replaced by $R_{ext}$ and $M_{ext}$ to calculate the breakout time from the envelope.  We have 
\begin{equation}
\label{tb_ext}
t_{b,ext} \simeq 120 C(\delta_{ext}) L_{j,46}^{-1/3} M_{ext,-3}^{1/3} R_{ext,13}^{2/3} \left(\dfrac{\theta_0}{0.2}\right)^{4/3} \text{\,s},
\end{equation}
where in this case we have scaled to $\delta_{ext} = 0$ via $C(\delta_{ext}) = \left[\frac{3}{5}\frac{5-\delta_{ext}}{3-\delta_{ext}}\right]^{2/3}$, which maximizes $t_{b,ext}$.  Combining equations (\ref{tb_star}) and (\ref{tb_ext}) with the parameters $M_c \approx 2\,M_\odot$, $M_{ext} \approx 4 \times 10^{-3}\,M_\odot$, and $R_{ext} \approx 9 \times 10^{12}$\,cm inferred in Section \ref{optical}, and an assumed core radius $R_c \sim 10^{11}$\,cm, we find the total breakout time
\begin{equation}
t_b \simeq 230 L_{j,46}^{-1/3} \left(\dfrac{\theta_0}{0.2}\right)^{4/3} \text{\,s}.
\end{equation} 
The condition $t_b < t_L$ is satisfied as long as $L_{j,46} \ga 5 \times 10^{-4} (\theta_0/0.2)^4$, which generally holds in our model so long as the jet is reasonably beamed.

To ensure that the interaction with the envelope is dominated by the supernova ejecta, the supernova energy should exceed the energy of the jet-blown cocoon, so that the former overtakes the latter.  The energy deposited into the cocoon up to breakout is $E_c \sim L_j (t_b - R_b/c)$ \citep{lazzati15}, where $R_b$ is the breakout radius.  There are two dynamically distinct cocoons that can potentially disturb the stellar envelope.  First, while the jet is within the stellar core, material entering the jet head escapes sideways to form a cocoon of shocked stellar matter.  When the jet breaks out of the stellar core and enters the surrounding envelope, this ``stellar cocoon" also breaks out and begins to sweep the envelope as it expands outwards.  Then, as the jet continues to propagate through the envelope, it blows a second cocoon containing shocked envelope material.  This ``envelope cocoon" expands laterally as the jet propagates, and then breaks out into the circumstellar medium once the jet reaches the envelope's edge.  Here, we show that these cocoons have a negligible effect on the envelope dynamics compared to the supernova, because the stellar cocoon is too slow and the envelope cocoon is too narrow.

Consider first the stellar cocoon.  While traversing the star, the jet head is nonrelativistic, so $t_b \gg R_b/c$ and essentially all of the energy goes into the cocoon, i.e. $E_{c,*} \sim L_j t_{b,*}$.  From equation (\ref{tb_star}), we have
\begin{equation}
\label{Ec}
E_{c,*} \sim 3 \times 10^{48} L_{j,46}^{2/3} M_{c,1}^{1/3} R_{c,11}^{2/3} \left(\dfrac{\theta_0}{0.2}\right)^{4/3} \text{\,ergs}.
\end{equation}
Here and for the rest of this section, we ignore order-unity factors that depend on $\delta_*$ or $\delta_{ext}$.  When $\beta_h < 1$, the cocoon expands sideways with speed $\beta_c \approx \beta_h \theta_0$, resulting in a cocoon opening angle $\theta_c \sim \beta_c/\beta_h \sim \theta_0$ \citep{bromberg}.  The mass entrained in the cocoon at the time of breakout is therefore
\begin{equation}
\label{Mc}
M_{c,*} \sim \dfrac{\theta_0^2}{2} M_c \sim 0.2 M_{c,1} \left(\dfrac{\theta_0}{0.2}\right)^{2}\,M_\odot.
\end{equation}
After breakout the cocoon material expands with typical speed $v_{c,*} \sim (2E_{c,*}/M_{c,*})^{1/2}$, so we have
\begin{equation}
v_{c,*} \sim 10^{8} L_{j,46}^{1/3} M_{c,1}^{-1/3} R_{c,11}^{1/3} \left(\dfrac{\theta_0}{0.2}\right)^{-1/3} \text{\,cm\,s}^{-1}.  
\end{equation}
This is generally much slower than $v_{ext} \approx 3 \times 10^9$\,cm\,s$^{-1}$ in our model, with $v_{c,*} < v_{ext}$ for $L_{j,46} \la 3 \times 10^4 M_{c,1} R_{c,11}^{-1} (\theta_0/0.2)$.  Therefore, the fast supernova ejecta rapidly overtake the stellar cocoon.

As discussed above, the jet stays collimated in the envelope, but the jet head may become relativistic.  In this limit, the lateral speed of the cocoon is $\beta_c \sim \tilde{L}^{1/2} \theta_0$ \citep{bromberg}, and since $\beta_h \approx 1$, $\theta_c \sim \beta_c$.  For a collimated jet, $\tilde{L} \la \theta_0^{-4/3}$ \citep{bromberg}, so we have $\theta_c \la \theta^{1/3} < 1$.  As the pressure of the envelope cocoon rapidly drops after it breaks out from the envelope's edge and expands freely into the low-density circumstellar medium, little sideways expansion through the envelope is expected after breakout.  Thus, as long as $\theta_0$ is small, the passage of the jet and envelope cocoon leaves the envelope relatively intact, and the SN-envelope interaction is quasi-spherical.  Note that it is not strictly necessary for the jet to be collimated by the envelope in our model.  In principle the jet may be uncollimated, with $\tilde{L}$ somewhat larger than $\theta_0^{-4/3}$, as long as $\theta_c$ remains small, but in practice this regime is not attained in GRB 060218.

Ideally, the jet head should break out of the stellar core before the SN shock.  This guarantees that the jet will reach the edge of the envelope before the SN, so the jet will be seen first.  Comparing the SN breakout time $t_{SN} \sim R_* (2E_{SN}/M_{SN})^{-1/2}$ to the breakout time in equation (\ref{tb_star}), one finds $t_{b,*}/t_{SN} \sim 0.6 L_{j,46}^{-1/3} R_{c,11}^{-1/3} \left(\frac{M_c}{2\,M_\odot}\right)^{-1/6} \left(\frac{E_{SN}}{2 \times 10^{51} \text{\,erg\,s}^{-1}}\right)^{1/2} (\theta_0/0.2)^{4/3}$.  This condition is satisfied for $L_{j,46} \ga 0.2 \left(\frac{M_c}{2\,M_\odot}\right)^{-1/2} \left(\frac{E_{SN}}{2 \times 10^{51} \text{\,erg\,s}^{-1}}\right)^{3/2} (\theta_0/0.2)^4$, which is only sometimes met for the parameters considered here.  However, even if the SN shock reaches the edge of the star first, the jet breaks out soon after.  This is because, after the SN crosses the core, the core density drops as $\rho_c \propto (v_{SN} t)^{-3}$, and as $\beta_h$ depends inversely on $\rho_c$, the jet soon accelerates to $c \beta_h > v_{SN}$.  Thus, it may be possible for this constraint to be violated, and we do not rule out models for which $v_{SN} > c \beta_h$ initially.

\begin{figure}
\includegraphics[width=\columnwidth]{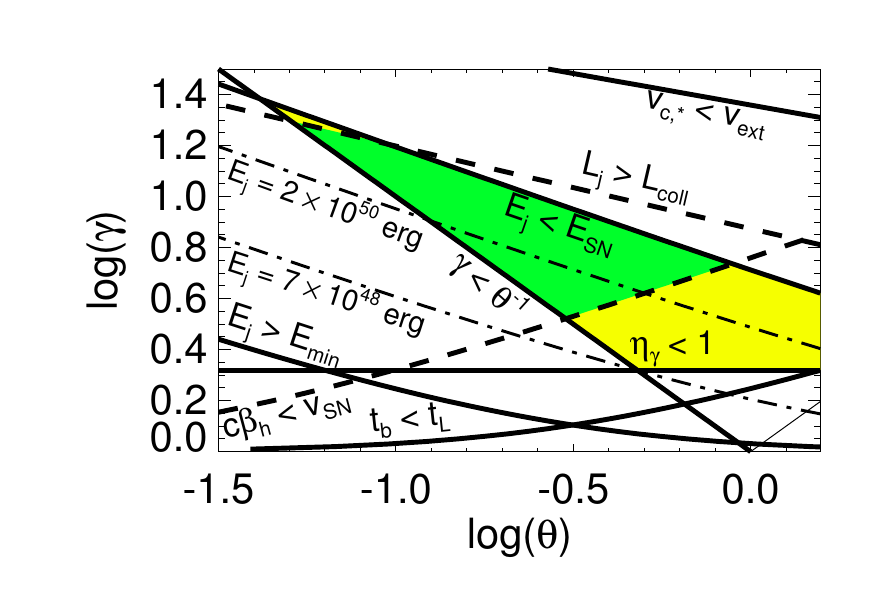}
\caption{Constraints described in Section \ref{jet_propagation}, depicted in the $\theta$--$\gamma$ plane.  The conditions $E_j > E_{min}$, $E_j > E_{SN}$, $\eta_\gamma < 1$, $\gamma \ga \theta^{-1}$, $t_b < t_L$, and $v_{c,*} < v_{ext}$ are drawn as heavy solid lines.  Two other constraints that may be marginally violated ($L_j < L_{coll}$ and $c \beta_h > v_{SN}$) are shown as dashed lines.  The regions of parameter space that satisfy all the constraints, and only the strict constraints, are painted green and yellow, respectively.  Curves of constant jet energy are shown as thin, dash-dotted lines.  See the text for discussion.}
\label{theta-gamma}
\end{figure}

We can get a grasp on the allowed region of $\gamma$--$\theta_0$ parameter space by using equations (\ref{Lj}) and (\ref{Ej}) to convert the conditions $\eta_\gamma < 1$, $E_j < E_{SN}$, $E_j \ga E_{min}$, $\gamma \ga \theta_0^{-1}$, $t_b \la t_L$, $v_{c,*} < v_{ext}$, and $c \beta_h < v_{SN}$ to relations between $\gamma$ and $\theta_0$.  We show the result in Figure \ref{theta-gamma}.  We see that the available parameter space is bound chiefly by $E_j < E_{SN}$ from above, $\gamma \ga \theta_0^{-1}$ from the left, $\theta_0 \la \pi/2$ from the right, and $\eta_\gamma \la 1$ from below.  The other important conditions are always satisfied when these four constraints are met.  Note that the conditions related to the jet and cocoon do not necessarily apply when $\theta_0$ is large, because in that limit the explosion is quasi-spherical instead of jet-like, but the conditions on the overall burst energetics are still relevant.  The possible values of $E_j$, $\gamma$, and $\theta_0$ lie in the range $7 \la E_{j,48} \la 2 \times 10^3$, $2 \la \gamma \la 25$, and $0.04 \la \theta_0 \la \pi/2$.

Three general classes of solution can satisfy all of the necessary conditions:
\begin{itemize}
\item \textbf{Low kinetic energy, narrow beam, low Lorentz factor:} 
For low jet energies, e.g. $E_j \simeq 7 \times 10^{48}$\,erg\,s$^{-1}$, the jet is confined to a narrow range around $\theta_0 \simeq 0.5$ and $\gamma \simeq 2.1$.  The isotropic jet energy in this case is $E_{iso} \simeq 6 \times 10^{49}$\,erg\,s$^{-1}$, and the radiative efficiency is $\eta_\gamma \simeq 0.5$, implying a mildly hot jet.  The kinetic energy during the afterglow phase is $E_k \simeq 3 \times 10^{48}$, which gives $-2.5 \la \log \epsilon_e \la -0.8$, $-3.6 \la \log\epsilon_B \la -0.5$, and $2.5 \la \log n \la 3.9$ using the bottom panel of Fig.~\ref{radioconstraints}.  This solution is similar to that of \citet{toma07}, who also inferred a mildly hot jet.
\item \textbf{High kinetic energy, narrow beam, high Lorentz factor:} 
For higher kinetic energies, the model is less restrictive: for example, $E_j = 2 \times 10^{50}$\,ergs gives $0.1 \la \theta_0 \la 1.5$, allowing for either narrow or wide jets.  In the narrow jet case of $\theta_0 \simeq 0.1$, we have $E_{iso} \simeq 4 \times 10^{52}$\,erg\,s$^{-1}$ and $\gamma \simeq 10$.  The radiative efficiency in this case is low, $\eta_\gamma \simeq 2 \times 10^{-3}$, and therefore $E_k \approx E_j$.  In order to accommodate the higher energy, this model requires lower-than-standard values for $\epsilon_e$ and/or $\epsilon_B$: $-5.3 \la \log \epsilon_e \la -2.6$ and $-5.4 \la \log\epsilon_B \la -0.5$.  For this reason, this scenario has not been considered previously.  The CSM density in this case is $3.5 \la \log n \la 5.7$.  
\item \textbf{High kinetic energy, wide beam, high Lorentz factor:} 
A high jet energy directed into a wide ($\theta_0 \rightarrow \pi/2$) outflow is also allowed.  In this case $E_j \approx E_{iso}$ .  For $E_j \simeq 2 \times 10^{50}$\,ergs, we find $\gamma \simeq 2.6$ and $\eta_\gamma \simeq 0.3$.  Using the top panel of Fig.~\ref{radioconstraints}, we find $-3.9 \la \log \epsilon_e \la -2.7$, $-3.1 \la \log\epsilon_B \la -0.5$, and $1.7 \la \log n \la 3.0$  This model is similar to the model proposed by \citet{fp06}, and also consistent with the picture in \citet{nakar15}, since in that case there is reason to expect a quasi-spherical explosion.
\end{itemize}
 
Our picture favors models with a narrow jet, because a wide jet would considerably disrupt the circumstellar envelope, which is problematic for our optical model discussed in Section \ref{optical}.  However, we do not have a strong reason to prefer a high jet energy versus a low energy one; each option offers some advantage.  The low energy case uses typical values for $\epsilon_e$ and $\epsilon_B$, and more readily transitions to a nonrelativistic outflow without the need for an extreme CSM density.  On the other hand, the high energy case is compatible with a wider variety of jets with various $\theta_0$ and $\gamma$, and requires a less efficient dissipation mechanism since $\eta_\gamma \ll 1$.

%************************************prompt nonthermal emission************************************************
\subsection{Prompt nonthermal emission}
\label{prompt_nonthermal}

Because GRB 060218 adheres to several well-known GRB correlations, it is worth considering whether this object has the same emission mechanism as standard GRBs.  In this section, we attempt to glean as much as possible about the emission mechanism in GRB 060218, assuming that the source of the nonthermal X-rays is a mildly relativistic jet.  To do so, we first lay out a simple empirical description for the prompt emission that preserves the essential features of the observed spectrum and light curve.  We then consider a number of jet scenarios for the prompt emission.

\subsubsection{A simple empirical description of GRB 060218}

\begin{figure}
\includegraphics[width=0.5\textwidth]{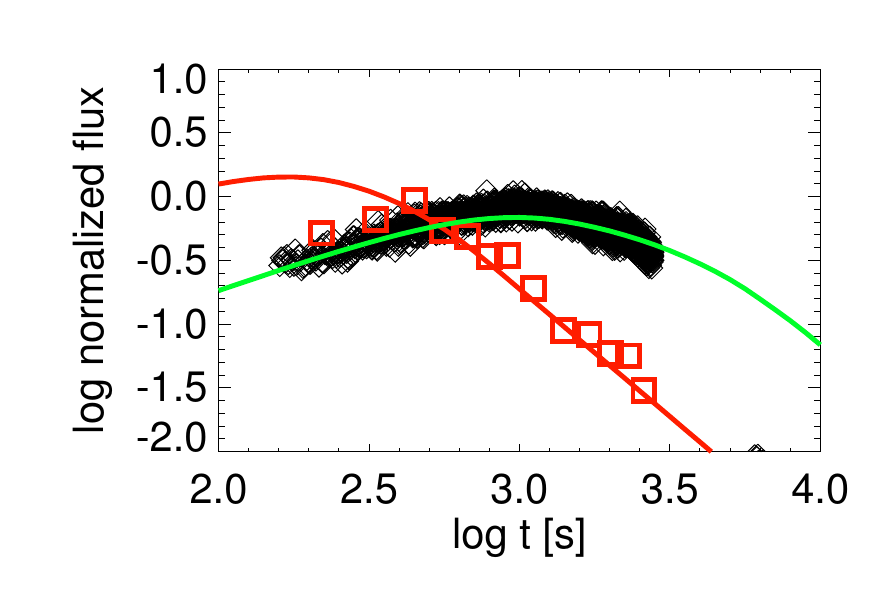}
\caption{Light curves obtained using a Band spectrum with time-varying parameters $E_p \propto t^{-1.6}$ and $F_0 \propto t^{1.0}$, and constant spectral indices $\beta_1 = -0.13$ and $\beta_2=-1.9$.  Model light curves in the 0.3$-$10 keV band (green) and the 15$-$150 keV band (red) are compared with XRT data from \protect\citet{campana06} (black diamonds) and BAT data from \protect\citet{toma07} (red squares).  Both light curves are fit well after 400 s, although the 15--150 keV flux is too high at early times.}
\label{bandmodel}
\end{figure}

The nonthermal emission observed in the XRT and BAT bands appear to have the same origin \citep{liang06}, and throughout the prompt phase, the joint BAT-XRT spectrum is best fit by a Band function \citep{toma07}.  Here, we aim to find the least complex model that fulfills these criteria while also fitting the light curves at various frequencies.  Consider a simple Band function model for the prompt nonthermal spectrum, where the photon spectral shape is given by \citep{band93}
\begin{equation}
\label{band_function}
\begin{array}{lr}
F(E) = F_0 (E/E_c)^{\beta_1} e^{-E/E_0}, & E \le E_c \\
F(E) = F_0 (E/E_c)^{\beta_2} e^{\beta_2-\beta_1}, & E > E_c
\end{array}
\end{equation}
The quantities $E_0$ and $E_c$ are related to the peak energy by $E_0 = E_p/(1+\beta_1)$ and $E_c = (\beta_1-\beta_2)E_0$.  Let the spectral indices be constant, and suppose the parameters $E_p$ and $F_0$ vary as power laws in time, i.e. 
\begin{equation}
\label{Ep_powerlaw}
E_p = E_n t^{\alpha_p}
\end{equation}
and
\begin{equation}
\label{F0_powerlaw}
F_0 = F_n t^{\alpha_0}
\end{equation}
where $E_n$ and $F_n$ are normalization constants. In the low and high energy limits, we have respectively $F(E \ll E_p) \propto t^{\alpha_0-\alpha_p \beta_1}$ and $F(E \gg E_p) \propto t^{\alpha_0-\alpha_p \beta_2}$.  Joint XRT and BAT observations fit with a Band model give $\alpha_p = -1.6$, $\beta_1 \approx -0.13$, and $F_{BAT} \sim F(E \gg E_p) \propto t^{-2.0}$ \citep{toma07}.  There is considerable uncertainty in $\beta_2$, so we do not fix its value, but rather try several values in the observed range.  Choosing $\beta_2 = -1.9$ and $\alpha_0 =1$, we have $F(E \ll E_p) \propto t^{0.8}$, giving a model that roughly reproduces the time behaviour and relative flux in the XRT and BAT bands (see Figure \ref{bandmodel}).  These choices are consistent with observations of the early XRT light curve, which show $F_{XRT} \propto t^{0.7}$.

\subsubsection{Towards an inverse Compton model}

The presence of a thermal component in the XRT band with a time-scale and luminosity comparable to the nonthermal component motivates consideration of an inverse Compton (IC) mechanism for the prompt emission.  Before investigating physical situations that might lead to strong IC upscattering, we first discuss some features generic to IC models for GRB 060218.  For now, we assume only that some hot electrons are present in the burst environment, but place no condition on the location of these electrons or the mechanism of their acceleration.

First, if the primary photon source is approximately monochromatic (as is the case for the observed constant-temperature thermal component), then the quickly decaying peak energy implies that the scattering electrons are rapidly decelerating in the comoving frame.  If the source frequency is $\nu_0 \sim k_B T_0$ in the observer frame, and the electrons dominating the radiation have comoving Lorentz factor $\gamma_p$, then the IC component will peak at $E_p \sim \gamma_p^2 k T_0$.  This implies $\gamma_p \propto t^{-0.8}$, since $E_p \propto t^{-1.6}$ \citep{toma07} and $T_0$ is approximately constant \citep{campana06}.  Suppose the electrons are accelerated to characteristic Lorentz factor $\gamma_m$, and can cool to Lorentz factor $\gamma_c$ in time $t$.  If cooling is slow (i.e., $\gamma_m > \gamma_c$), then $\gamma_p \sim \gamma_m$.  On the other hand, if cooling is rapid ($\gamma_c > \gamma_m$), then $\gamma_p \sim \gamma_c$ when $p < 3$, or $\gamma_p \sim \gamma_m$ when $p > 3$, because the $\nu F_\nu$ spectrum goes as $\nu^{(3-p)/2}$ between $\nu_m \equiv \gamma_m^2 k T_0$ and $\nu_c \equiv \gamma_c^2 k T_0$ \citep{spn98}.  $E_p$ evolves from $\sim 40$\,keV at first detection to $\sim 1$\,keV at $2000$\,s \citep{kaneko06}, and $k_B T_0 \simeq\,0.17$ keV for most times, implying that $\gamma_p$ varies from $\sim 15$ to $2$ throughout the prompt phase.

Second, the scattering medium is at most moderately optically thick, i.e. the electron scattering optical depth is $\tau_e \la 1$.  This follows from the observation of distinct thermal and nonthermal components, since when $\tau_e \gg 1$, essentially all of the photons undergo multiple scatterings, resulting in a single nonthermal peak.  A rough estimate of $\tau_e$ can be made by comparing the number of thermal and nonthermal photons.  The thermal component, with peak luminosity $L_{th} \sim 1 \times 10^{46}$\,erg\,s$^{-1}$ and time-scale $t_{th} \sim 3000 $\,s, carries $N_{th} \sim L_{th} t_{th}/kT_0 \sim 1 \times 10^{59}$ photons.  The nonthermal component has $L_{nt} \sim 3 \times 10^{46}$\,erg\,s$^{-1}$ at peak, duration $t_{nt} \sim 1000$\,s, and peak energy $E_p \sim 1$\,keV near maximum light, and therefore contains $N_{nt} \sim 2 \times 10^{58}$ photons.  This implies $\sim 1/6$ of thermal photons are scattered by electrons with $\gamma_e \simeq \gamma_p$, i.e. $\tau_e \sim 0.2$.  Note that $N_{th} \propto L_{th} t \propto t^{1.8}$ according to our model for thermal emission in Section \ref{prompt_thermal}, while $N_{nt} \propto F_0 t \propto t^2$ according to the Band function model described earlier in this section.  Hence $\tau_e \propto t^{0.2}$ is approximately constant in time in the simplest description.

Third, the nonthermal luminosity exceeds the thermal luminosity throughout most of the evolution.  This implies that the Compton parameter $y \sim \tau_e \langle \gamma_e^2 \rangle \ga 1$.  $\langle \gamma_e^2 \rangle$ represents the average gain per scattering.  Because $\tau_e \la 1$, we require $\langle \gamma_e^2\rangle \ga \tau_e^{-1} \ga 1$ to get $y \ga 1$, suggesting at least mildly relativistic electrons.  Since the total nonthermal luminosity is $L_{nt} \sim \tau_e \langle \gamma_e^2 \rangle L_{th}$, we can write $y \sim L_{nt}/L_{th}$.  Near maximum light, when the nonthermal component peaks in the XRT band, we can estimate $y$ directly: at 1000 s we have $L_{nt}/L_{th} \sim y \sim 6$, and at 3000 s we have $L_{nt} \approx L_{th}$ and $y \sim 1$.  In our simple Band function description, $y \propto E_p F_0/L_{th} \propto t^{-1.4}$, consistent with the values above.

Finally, the prompt spectrum holds information about the distribution of scattering electrons.  For electrons distributed as a power law in energy with index $p$, the spectral slope above $\nu_c$ and $\nu_m$ is given by $F_\nu \propto \nu^{-p/2}$.  The high energy spectrum in GRB 060218 has typical index $\beta_2 \approx -1.5$, but $\beta_2$ varies from $-3$ to $-1$, implying $p$ is in the range $2$--$6$ with typical value $p \approx 3$.  While most GRBs have $p$ values closer to 2, $p=3$ is not outside of the observed spread in $p$ values \citep{pk02}.

In summary, any IC model for GRB 060218 should be in the limit of moderately small scattering optical depth but appreciable energy gain, so that the thermal component carries most of the \textit{photons} (i.e., $N_{nt}/N_{th} \sim \tau_e \la 1$), but the nonthermal component carries most of the \textit{energy} (i.e., $E_{nt}/E_{th} \sim y \ga 1$).  $\tau_e$ most likely varies slowly in time, while $y$ and $\gamma_p$ decrease rapidly.  We now look at how well three different IC models satisfy these criteria.

\subsubsection{The photospheric model}

In the past several years, prompt thermal X-rays have been inferred from spectral fits in a number of GRBs, prompting the investigation of Comptonized photospheric models for the primary radiation \citep[for a recent review, see][and references therein]{peer15}.  In this picture, some of the bulk kinetic energy is dissipated into leptons within the jet, perhaps by internal shocks or magnetic effects.  Depending on the optical depth at which the dissipation occurs, different emergent spectra are possible.  Numerical simulations by \citet{chhotray15} have shown that, for leptons initially distributed as a power law $N(E) \propto E^{-p}$ (as typically assumed for shock heating), the observed spectrum takes the form of a thermal component with a high-energy power-law tail for relatively low dissipation optical depth $\tau_{diss} \sim 0.01-0.1$.  This type of spectrum is qualitatively similar to the observed spectrum of GRB 060218.  If the dissipation optical depth is decreasing with time, then one expects the peak energy to continuously decrease and the nonthermal component to become relatively weaker until only the thermal component remains.  This, too, is qualitatively consistent with GRB 060218, where the nonthermal component gradually fades, leaving a blackbody-dominated spectrum by $\sim 7000$\,s.  

However, the photospheric view is not without its problems.  For one, it is unclear how quickly $E_p$ decays.    In cosmological GRBs, the peak of the nonthermal component and the temperature of the thermal component are observed to be correlated, i.e. $E_p \propto T_0^{\alpha_T}$ with $\alpha_T$ typically $1$--$2$, and $T_0$ typically evolves with time \citep{burgess14}.  GRB 060218 does not obey this correlation, nor does it show evidence for an evolving temperature.   Nevertheless, the photospheric model provides a reasonable framework to interpret the early spectrum in GRB 060218, and is deserving of further attention.  More work is needed to understand subphotospheric emission, particularly in the low-luminosity and small Lorentz factor limit, and in the case where a dense envelope surrounds the progenitor star, but detailed photospheric modeling is beyond the scope of the the current paper.

\subsubsection{IC emission from external shocks in a long-lived jet}

A long-lived central engine not only serves as a possible source of strong photospheric emission, but also influences ejecta dynamics by driving shocks into the surrounding medium.  Here, we investigate IC scattering of prompt thermal photons from relativistic electrons in an engine-sustained reverse shock as a source for the observed prompt nonthermal emission.

First, we argue against the forward shock (FS) as the predominant IC emission site, because in this case obtaining a rapid decline of \textit{both} the peak energy and the high energy light curve is difficult.  Let the density of circumstellar material (CSM) as a function of radius be $\rho \propto r^{-\alpha}$.  A steep $E_p$ decline implies a rapid deceleration, which in turn suggests a flat CSM density profile so that the FS sweeps up mass more quickly.  Yet, the optical depth through the shocked region $\tau \sim \kappa \rho r \propto r^{1-\alpha}$ actually \textit{increases} with time when $\alpha < 1$.  Compounding this with the rising thermal luminosity $L_{th} \propto t^k$, the peak spectral luminosity ($L_{\nu,max} \propto L_{th} \tau \propto t^k r^{1-\alpha}$ in the optically thin case) rises sharply with time in a flat density distribution.  Since the BAT band lies above the peak energy, the luminosity there scales as $L_{BAT} \propto L_{\nu,max} \nu_c^{1/2} \nu_m^{(p-1)/2}$ \citep{spn98}.  Whether the peak is due to $\nu_c$ or $\nu_m$, the rising $L_{\nu,max}$ makes it difficult to ever obtain a BAT flux that declines faster than the peak energy in the case where the FS dominates emission.

This problem is alleviated by considering the reverse shock (RS) as the emission site instead.  An immediate question is how to attain a declining peak energy at the RS, since generally this shock would go to higher Lorentz factor (in the fluid frame) as the flow is decelerated.  As it turns out, a long-lived central engine can help in this regard.  In the limit where the engine deposits mass into the RS more quickly than the FS sweeps mass from the CSM, the dynamics differ markedly from the typical GRB case.  To illustrate this, consider ejecta with mass $M_{ej}$ and Lorentz factor $\gamma_{ej}$ interacting with a cold CSM.  Previous authors \citep[e.g.][]{sp95} have shown that the outflow undergoes a dynamical transition when the mass swept by the forward shock, $M_{sw}$, becomes equal to $\sim M_{ej}/\gamma_{ej}$.  When $M_{sw} \ll M_{ej}/\gamma_{ej}$ the shocked shell at the ejecta-CSM interface coasts with a constant Lorentz factor $\gamma_{sh} \approx \gamma_{ej}$, and when $M_{sw}\gg M_{ej}/\gamma_{ej}$, $\gamma_{sh}$ evolves with time.  In the typical GRB case, where there is no continued mass or energy input from the central engine, $M_{ej}$ is constant while $M_{sw}$ grows over time; the system begins in the coasting state, and the shell starts to decelerate once $M_{sw}$ becomes sufficiently large.  However, for continuous mass ejection, $M_{ej} \propto \int \dot{M}_j dt$ also increases with time.  The dynamics will be altered if $M_{ej}$ grows faster than $M_{sw}$, which is possible for a steep CSM density gradient.  In that limit $M_{sw} > M_{ej}/\gamma_{ej}$ initially, and the shell \textit{accelerates}, eventually reaching a terminal Lorentz factor $\gamma_{sh} \approx \gamma_{ej}$ once $M_{sw} \ll M_{ej}/\gamma_{ej}$.  In this scenario, the RS decelerates steadily in the contact discontinuity frame.  Therefore, $E_p \propto \nu_m$ falls off quickly if the emission comes from a rapid-cooling reverse shock.  In a steep density gradient, $\tau \propto r^{1-\alpha}$ also decreases with time, making it possible for $L_{BAT}$ to decline faster than $E_p$ if the emission comes from the RS.

While this rough example serves to illustrate the benefit of prolonged engine activity, calculating the XRT and BAT light curves requires a more thorough model.  For an engine-driven outflow the resulting forward-reverse shock system can be divided into four regions -- the cold CSM, the shocked CSM, the shocked outflow, and the unchecked outflow -- which we label with subscripts 1 to 4, respectively.  Regions 2 and 3 are separated by a contact discontinuity.  Taking a constant adiabatic index of $4/3$, and assuming that the bulk Lorentz factor and internal energy in regions 2 and 3 do not change much between the shock and contact discontinuity, the dynamics of this system are governed by the equations \citep{sp95}:
\begin{equation}
\label{fdefine}
f \equiv \bar{\rho}_4(R_{rs})/\rho_1(R_{fs}),
\end{equation}
\begin{equation}
\label{fgeneral}
f = \dfrac{(4\gamma_2+3)(\gamma_2-1)}{(4\bar{\gamma}_3+3)(\bar{\gamma}_3 -1)},
\end{equation}
\begin{equation}
\label{gamma3general}
\bar{\gamma}_3 = \gamma \gamma_2 (1-\beta \beta_2),
\end{equation}
and
\begin{equation}
\label{radius}
R_{fs} \approx R_{rs} \approx \dfrac{\beta_2 ct}{1-\beta_2},
\end{equation}
where $R_{fs}$ is the radius of the FS and $R_{rs}$ is the radius of the RS.  These equations are valid in both the ultrarelativistic and mildly relativistic limit, but break down as the forward shock becomes nonrelativistic since in that case $R_{rs} \ll R_{fs}$.

To close the system of equations (\ref{fdefine})--(\ref{radius}), expressions for the densities $\rho_1$ and $\bar{\rho}_4$ as functions of radius $r$ are needed.  The outer wind density we parametrize by 
\begin{equation}
\label{rho1}
\rho_1 = 5 \times 10^{11} A_* r^{-2} (r/R_{ext})^{2-\alpha} \text{\,g\,cm}^{-3}.
\end{equation}
Here, $A_*$ is a parameter scaled to a pre-explosion mass-loss rate of $\dot{M}_{wind} = 10^{-5}\,M_\odot$\,yr$^{-1}$ and velocity $v_{wind} = 1000$\,km\,s$^{-1}$.  While $A_* =1$ is typical for Wolf-Rayet progenitors, a star with an extended envelope could have a different mass loss history.  $\alpha$ determines the slope of the CSM density profile, with $\alpha=2$ corresponding to the usual wind profile.  For a given $A_*$, winds with different $\alpha$ are scaled to the same density at $R_{ext}$.  The density of the inner wind is given by 
\begin{equation}
\label{rho4}
\bar{\rho}_4 = \dfrac{1}{4 \pi r^2 \beta \gamma} \dot{M}_{iso}(t_{emit}),
\end{equation}
with $\dot{M}_{iso}$ given by equation (\ref{Mdot}).  The time $t_{emit} =(\beta-\beta_2)t/\beta(1-\beta_2)$ takes into account the delay between the arrival of photons and matter from the central engine at the reverse shock.  Note that our model is only valid when $\alpha < 3$ (i.e., when the swept CSM mass is not negligible) and $k > -2$ (i.e., when the energy input from the central engine is not negligible).

We solved numerically the system of equations (\ref{fdefine})--(\ref{rho4}) to determine the dynamical variables.  For reference, we also present analytical solutions of these equations in the limit $\gamma \gg 1$ in Appendix \ref{appendix_a}.  Once $\gamma_2$, $\bar{\gamma}_3$ (or $\bar{\beta}_3$), and R are known, the spectral parameters $\nu_m$, $\nu_c$, and $L_{\nu,max}$ for the forward- and reverse-shocked regions can be determined using the standard theory, as outlined in Appendix \ref{appendix_b}, where we give analytical expressions for the maximum spectral power $L_{\nu,max}$  and break frequencies $\nu_c$ and $\nu_m$.  In calculating the spectrum, we approximate the thermal photons as monochromatic with frequency $\sim 3 kT_0$.  We take the blackbody temperature to be $0.14$\,keV as in \citet{kaneko06}, as we find this gives a better fit than the higher value inferred by \citet{campana06}.  The nonthermal spectrum is taken to have a power-law form with breaks at $\nu_m$ and $\nu_c$, and the BAT and XRT light curves are determined by integrating the spectrum over $15$--$150$\,keV and $0.3$--$10$\,keV, respectively.  Our best fitting models have $p>3$, so in all cases discussed here $E_p \simeq h\nu_m$.  The free parameters of our models are $\alpha$, $p$, $A_*$, $\gamma$, and the fractions of postshock energy $\epsilon_{e2}$ and $\epsilon_{e3}$ going into relativistic electrons in the FS and RS respectively.  We fix $k$, $L_0$, and $\xi$ to the values inferred from the observed thermal component, as in Section \ref{prompt_thermal}.

We calculated $L_{BAT}$, $L_{XRT}$, and $E_p$ over a range of parameter space.  In order to match the observed slope of $E_p$ and $L_{BAT}$, we fixed $\alpha$ to 2.7; for other values of $\alpha$, the fit for these observables is typically worse, although the fit to $L_{XRT}$ is sometimes improved.  We varied $p$ from 3.0--4.0 in steps of 0.1, $\log A_*$ from 2.0--5.0 in steps of 0.0125, $\gamma_e$ from 6.0--12.0 in steps of 0.025, and $\log \epsilon_{e3}$ from -3.5 to -1.5 in steps of 0.025.  In no case were we able to fit the spectrum reasonably with FS emission, so rather than vary $\epsilon_{e2}$, we fit the light curves with the sum of thermal emission and nonthermal RS emission, and then calculate an upper limit on $\epsilon_{e2}$ by assuming the FS contributes less than 30\% of the flux in XRT and BAT.  In addition, we place an upper limit on $\epsilon_{B}$ by demanding that the comoving energy density ($u_{rad}$) of thermal radiation at the RS be higher than the comoving energy density in magnetic fields ($u_B$).  For each model, we calculated the reduced chi-squared $\chi^2$ via a joint fit to the observed XRT luminosity from \citet{campana06} and the BAT luminosity and peak energy measured by \citet{toma07}.  We do not include the XRT data after $4000$\,s in the fit.  

\begin{figure}
\centering
\begin{tabular}{cc}
\includegraphics[width=0.5\textwidth]{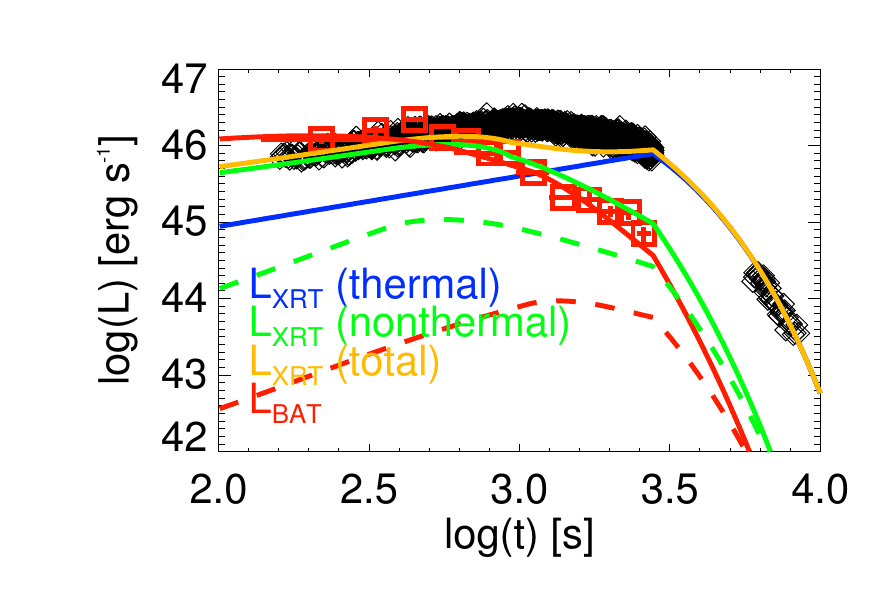} \\
\includegraphics[width=0.5\textwidth]{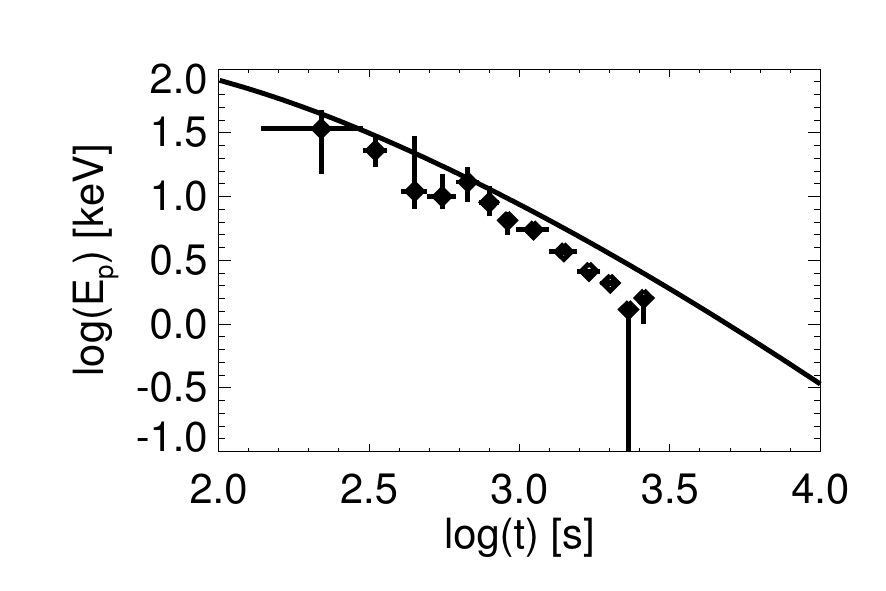}
\end{tabular}
\caption{ \textit{Top:}  XRT (0.3$-$10 keV) and BAT (15$-$150 keV) light curves for our reverse shock IC model.  When fixing $k=0.66$, $\xi=1$, $L_{0,46}=1$, and $\alpha=2.7$, the best-fitting parameters are $\epsilon_{e3} = 6.0 \times 10^{-3}$, $\gamma=10.8$, $A_* = 4900$, and $p=3.8$  The thermal, nonthermal, and total XRT luminosities are shown in blue, green, and orange respectively, while the BAT luminosity is drawn in red.  Dashed curves show the contribution of the forward shock to the BAT (red) and XRT (green) light curves, assuming $\epsilon_{e2}/\epsilon_{e3} = 0.05$. The black diamonds are XRT data from \protect\citet{campana06}, and the red squares are BAT data from \protect\citet{toma07}. \textit{Bottom:} The peak energy in our model, as compared to data from \protect\citet{toma07}. }
\label{ICmodel}
\end{figure}

  In some cases -- particularly when $A_*$ is rather large -- the optical depth of the shocked regions exceeds unity.  Our model, which assumes single scattering, is not valid in that case.  Thus, we discard models with high optical depth, keeping only models that become optically thin prior to $300$\,s.  After that cut is applied, our best-fitting model has $p=3.8$, $A_*  = 4900$, $\gamma = 10.8$, and $ \epsilon_{e3} = 6.0 \times 10^{-3}$.  This model is shown in Figure \ref{ICmodel}.   We find that the FS does not contribute substantially to the emission when $\epsilon_{e2} \la 0.05 \epsilon_{e3}$.  $u_{rad}/u_B > 1$ at all times in this model if $\epsilon_B < 2 \times 10^{-5}$.  If we relax the condition to $u_{rad}/u_B > 1$ only after 300 s, then $\epsilon_B \la 10^{-4}$.  While the low upper limit on $\epsilon_B$ is somewhat troubling, we note that other authors \citep[e.g.,][]{fp06,duran14} have also found a low value of $\epsilon_B$ compared to $\epsilon_e$ in GRB 060218.

While this model can plausibly fit the light curves, it cannot explain the low-frequency spectral shape: for $\nu_c < \nu < \nu_m$, we have $F_\nu \propto \nu^{-1/2}$, steeper than the observed spectrum $F_\nu \propto \nu^{-0.1}$.  However, we note that \citet{toma07} found a different spectral shape at low energies, $F_\nu \propto \nu^{-0.4}$, when using a cut-off power-law to fit the data instead of a Band function, so the observed $\beta_1$ seems to depend in part on the assumed spectral model.  We also find a high value of $p\simeq3.8$ that, while roughly consistent with the observed value of \citet{toma07}, is large compared to the value in typical GRBs \citep{pk02}.  An additional issue with our model is that it slightly underpredicts the XRT flux near peak by up to a factor of 2, and slightly overestimates the peak energy.  Furthermore, our model is only one-dimensional, and it does not take into any effect of collimation or sideways expansion of the jet.

Despite these issues and the crudeness of the model, the reverse shock IC interpretation does a reasonable job of capturing the basic behaviour of the light curves, and it has some attractive features.  Notably, of all the models we consider, this model is the only one that provides a natural explanation for the steep decline of $E_p$ and $L_{BAT}$.  Additionally, this type of emission is expected when a long-lived, dissipative jet is present, and should therefore contribute to the emission on some level (although, IC emission from the RS only dominates the contribution from external shocks under certain circumstances, as described above).  However, if other lepton populations (e.g., those excited by internal shocks or other dissipation in the jet interior) also strongly contribute to the emission, the external shock emission may not be observed.

Having an independent estimate for the outflow Lorentz factor allows us to break the degeneracy of our radio model discussed in Sections~\ref{radio_afterglow} and \ref{jet_propagation}, by calculating $E_{iso}$ directly.  Assuming $t_L$ corresponds to the turn-off time of the engine, we find $M_{iso} = 4.9 \times 10^{-3}\,M_\odot$ and $E_{iso} = 8.7 \times 10^{52}$\,ergs via equations (\ref{Mj}) and (\ref{Ej}).  Thus, the high energy, high Lorentz factor radio model is preferred.  The low upper limits on $\epsilon_B$ and $\epsilon_e$ in this section are consistent with the ranges inferred from the radio (see discussion at the end of Section~\ref{jet_propagation}).   The jet is cold and radiates inefficiently, with $\eta_\gamma = 6.9 \times 10^{-4}$.  Applying $\theta_0 \ga \gamma^{-1}$, we find that the true jet energy is $E_j \ga 4 \times 10^{50}$\,ergs. These results have interesting implications when compared to standard GRBs.  $E_{iso}$ and $E_j$ fall within the range typical for cosmological bursts \citep{piran04}, suggesting that the total kinetic energy released in GRB 060218 is not unusual, although it is released over a longer time.  The main factor that distinguishes GRB 060218 from the bulk of observed GRBs is therefore its radiative efficiency: whereas most bursts have $E_{\gamma,iso} \approx E_{iso}$ \citep{piran04}, our model for GRB 060218 predicts $E_{\gamma,iso} \ll E_{iso}$.  This fact is closely linked to the low values we deduced for the $\epsilon$ parameters, which may perhaps be related to the lower bulk Lorentz factor or the long engine lifetime.  For now, this is only speculation, but the possibility that standard GRBs are a corner case where the radiative efficiency is high (due, perhaps, to a higher Lorentz factor or a more abrupt deposition of kinetic energy), while most collapsar events go unobserved because of a much lower radiative efficiency, is intriguing. 

In addition, we can deduce some properties of the CSM near the progenitor from the inferred value of $A_*$.  If it extends in to $R_{ext}$, the wind is optically thick to electron scattering, with total optical depth $\tau_w = \int_{R_{ext}}^{\infty} \rho_1 \kappa dr \approx 30$.  However, because of the steep density gradient, the high optical depth is due mostly to material very close to $R_{ext}$.  In fact, the shock radius in our model is $R \approx 5 \times 10^{13}$ cm at the time of first observation; our model does not constrain the wind density at radii less than this.  The optical depth of the wind is small compared to the envelope optical depth, so the addition of such a wind does not affect the optical model discussed in Section~\ref{optical}.  This wind cannot be the origin of the high value of $N_H$, however, as the absorbing column through the wind changes as the shock propagates outward, while the observed $N_H$ is constant.  

The extent of the wind is not known, but the equipartition radius $R_N = 1.3 \times 10^{16}$ cm at day 5 gives an upper limit, since as discussed in Section \ref{radio_afterglow} the radio observations imply a constant density CSM.  The total mass of the wind is therefore $M_w < \int_{R_{ext}}^{R_N} 4 \pi r^2 \rho_1 dr = 3.6 \times 10^{-3} M_\odot$.  This is comparable to the isotropic mass of the jet, so it is possible that the jet undergoes some deceleration while sweeping the outer layers of the envelope.  In addition, even though the terminal Lorentz factor is $\gamma_2 \approx \gamma = 10.8$ in our model, the transition to the coasting state is quite gradual: we find $\gamma_2$ only reaches $\approx 5$ by the end of the prompt phase at $t_L$.  It therefore seems plausible that the jet could decelerate to $\beta\gamma \sim 1$ by day 5, as implied by radio observations.  We stress, however, that there is still some tension in producing the flat radio light curve with a mildly relativistic outflow.

After the source of thermal photons fades away, the emission from external shocks will be dominated by synchrotron radiation.      Since the overall SED appears incompatible with a single synchrotron spectrum, this component should not overwhelm the optical emission and dust echo afterglow emission observed at the same time.  Since the jet begins to decelerate shortly after $t_L$ in our model, the synchrotron emission peaks near $t_L$.  At that time we find that the critical synchrotron frequencies are $\nu_m \simeq 6 \times 10^{-9}$\,Hz and $\nu_c \simeq 4 \times 10^{12}$\,Hz; both are far below the optical band.  We calculate a peak synchrotron $\nu F_\nu$ flux of $5 \times 10^{-15}$\,erg\,cm$^{-2}$\,s$^{-1}$ $2 \times 10^{-17}$\,erg\,cm$^{-2}$\,s$^{-1}$, respectively at $5 \times 10^{14}$\,Hz and $1$\,keV.  This is far below the observed $\nu F_\nu$, which is $\sim 10^{-11}$\,erg\,cm$^{-2}$\,s$^{-1}$ in both the X-ray and optical bands \citet{toma07}.  The main reason synchrotron emission from the jet is so weak compared to cooling envelope emission and dust scattering of the prompt light is the low values of $\epsilon_e$ and $\epsilon_B$.  This answers the question posed at the end of Section~\ref{xray_afterglow}, suggesting that the primary reason a dust echo is observed in GRB 060218 is because of low values of the microphysical parameters.
 
\subsubsection{Other models for jet emission}

Here we briefly consider some other possibilities for the prompt emission, but each is problematic.  Thus, we prefer an inverse Compton interpretation for the prompt emission in GRB 060218.

Synchrotron emission from external and/or internal shocks is also expected for relativistic jets.  The standard FS synchrotron model, with constant external density and no continuous energy injection (i.e., $\alpha=0$ and $k=-2$), gives $E_p \propto \nu_m \propto t^{-3/2}$ and $F(E \gg E_p) \propto t^{(2-3p)/4}$ in the rapid cooling limit \citep{spn98}.  When $p\simeq 3$, this gives a time behaviour similar to the observed one.  However, the peak energy in this case is much too low to explain observations, even for unphysically high explosion energies.   

Internal shock models, in which ejecta shocked by the collision of successive engine-launched shells radiate via synchrotron, remain a prominent model for the prompt radiation in cosmological GRBs \citep{piran04}.  This picture provides a natural interpretation of the high degree of variability in GRB light curves, as the many shell collisions give rise to multiple peaks.  GRB 060218, with its smooth, single-peaked light curve, may therefore be hard to explain in an internal shock context, unless an additional mechanism acts to smooth out the light curve.  It is unclear, as well, how the presence of an extended envelope could affect the internal shock signature.

In some cases, steep decays in the prompt GRB light curve have been attributed to a kinematical effect, wherein the observer continues to see emission from high-latitude parts of the curved emission region after the prompt emission process ends.  This phenomenon, known as the curvature effect, leads to fainter and softer emission over time because of relativistic beaming.  \citet{toma07} already investigated curvature effects for GRB 060218, and showed that the simultaneous steep decay of the peak energy and high energy light curve is inconsistent with this interpretation.

%-------------------------------------------------------------------------------------------------------------------------------
%***********************************************SECTION 5**************************************************
%---------------------------------------------------------------------------------------------------------------------------

\section{Discussion}
\label{discussion}

We have presented a model for the peculiar GRB 060218 in which the prompt X-ray emission arises from a low-power jet and the early optical emission is powered by fast SN ejecta interacting with a low-mass circumstellar envelope.  Our picture has some features in common with the recent model of \citet{nakar15}, where the prompt X-ray and optical emission is produced by a choked jet interacting with a circumstellar envelope.  In both cases, a jet is needed to decouple the mildly relativistic outflow from the SN, and an extended envelope of similar mass ($\sim 0.01$ $M_\odot$) and radius ($\sim 100$ $R_\odot$) is inferred.  Both models provide a reasonable explanation for the radio afterglow flux, although Nakar's model has an advantage in explaining the shallow slope of the light curve.  Neither model can account for the X-ray afterglow through external shock synchrotron radiation alone.  There are several key differences between the models, however.  We differ on the jet properties (we suggest a low-power, long-duration jet, whereas Nakar uses a more typical GRB jet), the origin of the prompt X-rays (we prefer a dissipative jet and some Compton scattering process, whereas Nakar posits shock breakout), and the power source for the cooling envelope emission seen in the optical band (we suggest that it is driven by the underlying SN, whereas Nakar proposes a smothered jet explosion).  A detailed discussion of the strengths and weaknesses of each model is therefore warranted.

An advantage of Nakar's model is that the luminosity and time-scale of the jet take on typical GRB values.  In addition, shock interaction naturally produces a smooth, single peaked light curve in X-rays, as observed \citep{ns12}.  This model also helps to explain the lack of a jet break in the radio, since the jet outflow becomes quasi-spherical before leaving the envelope.  A wind-like CSM profile is also inferred for afterglows powered by shock breakout \citep{duran14}, which is expected for a WR star progenitor.  On the other hand, the high value ($\sim 50$ keV) and slow decay ($\propto t^{-(0.5-1)}$) of the prompt peak energy that one expects in the shock breakout scenario \citep{ns12} seem hard to reconcile with direct measurements of the peak energy that show it declines steeply as $t^{-1.6}$ and with a value of $\sim$ a few keV throughout most of the prompt phase \citep{toma07}.  The fact that the prompt optical and X-ray emission are observed simultaneously, and that they each evolve smoothly from the earliest observation, also seems hard to interpret in a shock interaction model where the observed radiation evolves from a nonequilibrium state toward thermal equilibrium.  A better understanding of the expected X-ray signal from mildly relativistic shocks in low-mass envelopes is needed to determine whether these problems can be resolved. The origin of the prompt thermal X-ray component is also unclear in the choked jet model, since the inferred photospheric radius is considerably smaller than the envelope radius.  Finally, as discussed above, there is some question of whether an ultrarelativistic jet can truly be sphericized in a 0.01 $M_\odot$ envelope.  Detailed hydrodynamical simulations will be crucial to fully understand how such an envelope affects jet propagation and sideways expansion.

Our low-luminosity jet model comes with its own merits and drawbacks.  A jet origin for the prompt X-rays and $\gamma$-rays places GRB 060218 at the low-luminosity, long-duration end of a continuum of GRB processes.  In this unified picture, similarities to cosmological bursts (such as satisfying the Amati and lag-luminosity correlations) are perhaps not surprising, although as in typical GRBs the physical origin of these correlations is not well understood.  None the less, these coincidences are not easily accounted for in the shock interaction view.  Furthermore, the presence of a thermal component is expected for a dissipative jet, and decoupling the prompt X-ray and optical emission removes problems with the X-ray to optical evolution.  However, the low-power jet interpretation inherits one usual problem with jet models, namely that the prompt emission mechanism in relativistic jets is still not well understood.    Also, since a low-luminosity jet stays collimated while it drills through the circumstellar envelope, our model requires the jet to become nonrelativistic in the CSM, which may be difficult unless some additional mass close to the star helps to decelerate the jet.  While we infer a wind-like CSM at small radii where the prompt X-rays are emitted, we find that a uniform circumburst density is needed beyond $ 10^{16}$\,cm where the radio is emitted.  This is contrary to usual expectations for a WR progenitor.  Finally, we require an unusually long-lived, low-power central engine, the origin of which is unclear.  

This last point deserves more discussion.  A shortcoming of our model, as with many engine driven models, is the need to prescribe unknown properties of the central engine.  A simple parametrization glosses over many of the finer details of compact object formation and jet launching, the physics of which are not yet fully understood.  In particular, producing a long-duration, low-luminosity engine from a nascent black hole presents problems: \citet{aloy05} have shown in their black hole simulations that a minimum jet luminosity of $\sim 10^{49} \text{ erg s}^{-1}$ is needed to overcome the ram pressure of accreting material, and black hole-driven engines tend to operate on time-scales much shorter than $\sim 10^3$ seconds.  However, the SN might clear away infalling material, thus allowing a lower luminosity jet to propagate.  It is unclear whether the formation of a neutron star (or magnetar) could drive the type of mildly relativistic outflow we require, but the longer time-scales involved are more consistent with the long-lasting prompt emission observed \citep[for further discussion, see][]{toma07}.    Magnetar-powered scenarios are particularly intriguing in light of the recent result of \citet{greiner15}, who claim the detection of magnetar-driven superluminous SN associated with the ultra-long GRB 111209A.    Here, we only aim to show that a low-luminosity outflow, if present, can explain many features of the prompt thermal and nonthermal emission.  Note that several other bursts, such as GRB 130925A \citep{evans14} and the ultra-long bursts discussed by \citet{levan13}, may also require long-lived central engines, so this problem is not unique to GRB 060218.

The lack of variability in the light curves of GRB 060218 and GRB 100316D merits further investigation, as well.  If the typical GRB variability originates from relativistic turbulence, then the smooth light curves observed in some LLGRBs could be ascribed to the lack of highly relativistic material \citep{nk09}.  Even in the absence of relativistic effects, a light jet lifting heavier external material would give rise to Rayleigh-Taylor instabilities that may induce light curve fluctuations.  This could be circumvented by, e.g., a Poynting-flux dominated jet, but a matter-dominated jet is required to produce the prompt nonthermal X-rays through IC processes.  The smooth light curve also constrains the degree of clumpiness which can be present in the CSM.  Detailed numerical simulations will be needed to characterize the amount of variability expected from a mildly relativistic jet as it penetrates the star and envelope, breaks out from the envelope's edge, and sweeps the surrounding CSM.

We note that, because our model involves an on-axis jet, we cannot appeal to geometric effects to increase the rate of 060218-like events.  However, our model \textit{does} imply a unique, non-standard progenitor different from the usual high-mass WR stars thought to give rise to most cosmological LGRBs.  Thus, our explanation for the high rate of LLGRBs is simply that LLGRB progenitors are intrinsically $10$--$100$ times more common.  Assuming that the presence of a long-lived, low-luminosity jet is also somehow tied to the progenitor structure, such jets might also be more common than ultrarelativistic, short-lived ones, but we are biased against observing them due to their low power.  In the model of \citet{nakar15}, the progenitor is again different from the standard one, but the prompt emission is roughly isotropic as well, so that the higher event rate is due to some combination of geometric and intrinsic effects.

\begin{figure}
\includegraphics[width=\columnwidth]{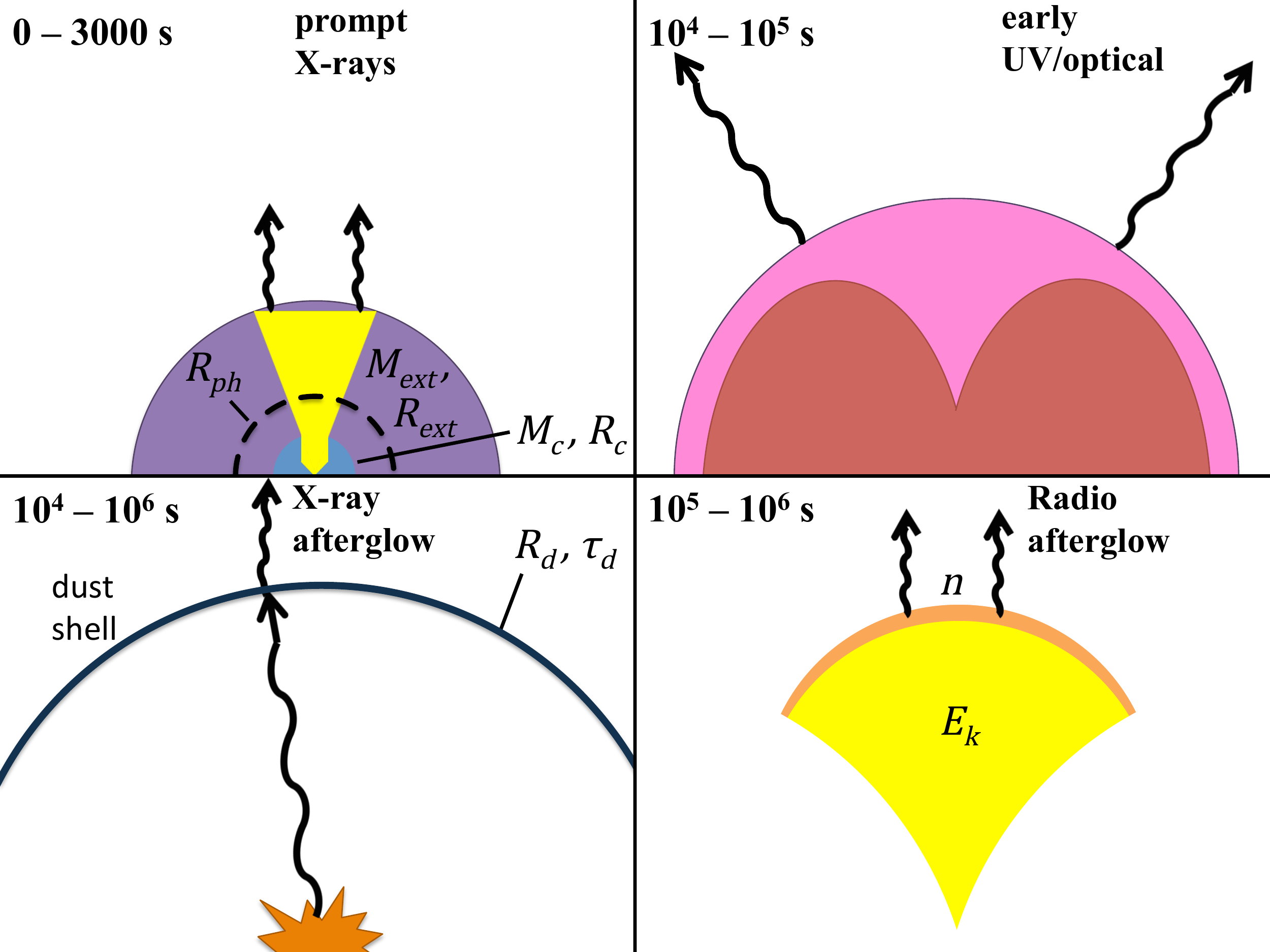}
\caption{ A hybrid model for the prompt emission.  \textit{Upper left: } A jet is launched with short time-scale compared to the envelope breakout time, as in \protect\citet{nakar15}. However, the jet does not have time to become quasispherical before breaking out; it undergoes significant deceleration, and possibly a small degree of lateral spreading, but the explosion breaks out primarily in the forward direction, leaving the envelope mostly intact.  Thermal emission could be observed, e.g. from the walls of the jet cavity, once material clears out along the line of sight.  \textit{Upper right: }  As in our model, the fast SN ejecta heat the remaining envelope, which cools through optical radiation.   \textit{Lower left: } The X-ray afterglow is produced from dust scattering, as described in Section \ref{xray_afterglow}.  \textit{Lower right: } The radio afterglow comes from a nonrelativistic, quasispherical blast wave.  Because the ejecta are already decelerated to $\beta \gamma \sim 1$ by the envelope, a spherical flow is more readily achieved than in our jet breakout model. }
\label{cartoon2}
\end{figure}

It may be possible to construct a "hybrid" model that retains some of the best features of both our model and Nakar's.  This speculative model is depicted in Figure \ref{cartoon2}.  Suppose that the central engine switches off while the jet is traversing the envelope (as in Nakar's model), but let the envelope mass be smaller (as in our model) so that the outflow decelerates significantly and expands sideways somewhat, but does not have time to sphericize before breaking out.  The explosion then breaks out aspherically, with shock breakout emission expected from near the poles.  If dissipation continues to occur after the cessation of engine activity, a thermal component might also be observable once the ejecta clear out.  After breakout, the ejecta expand into the low-density CSM, eventually producing the radio synchrotron emission as electrons are accelerated by the external shocks.  Since the outflow expands preferentially into the CSM post-breakout, a quasi-toroidal envelope remnant is left behind, which is shocked by the fast SN ejecta and then emits the prompt optical emission as it cools.  The X-ray afterglow is produced by dust scattering, as in Section~\ref{xray_afterglow}.  As with our model, this scenario gives a possible explanation for the thermal emission, and since the optical and X-ray are decoupled there are no concerns with the X-ray to optical spectral evolution.  Yet, as with Nakar's model, this case generates a smooth prompt X-ray light curve via shock breakout, and more easily explains the radio because the initial outflow is wider than a jet.  However, the expected signal from an aspherical shock breakout in the relativistic limit has not been calculated in detail, which is an important caveat.

Moving past the prompt emission, we have also shown that a dust echo model gives a reasonable fit to the X-ray afterglow light curve and spectral index evolution.  The dust echo model used only an empirical fit to the prompt light curve and spectrum, and therefore is insensitive to the mechanism of prompt emission.  Moreover, the scattering angle from the dust grains, $\theta_d \approx (2ct/R_d)^{1/2} \sim 1\degree (R_d/30 \text{ pc})^{-1/2} (t/10 \text{ days})^{1/2}$, is small, so the echo emission depends only on the prompt radiation roughly along the observer's line of sight.  Thus, the dust echo interpretation applies equally well whether the prompt emission originates from a low-luminosity jet or from shocked gas.   If the reason for the small radiated energy in bursts like GRB 060218 is small values of $\epsilon_e$ and $\epsilon_B$, as our model suggests, then dust echo type afterglows should commonly accompany this class of bursts, because the synchrotron emission from external shocks will be weak.  So far, this is borne out by observations, as the afterglow GRB 100316D is also consistent with a dust echo.  Why the synchrotron efficiency is poor, and whether this is related somehow to the long burst duration, remains to be worked out.  

If our picture for GRB 060218 is correct, one would expect to observe broad-lined Type Ic SNe with accompanying mildly relativistic radio afterglows, but without a prompt X-ray component, when viewing GRB 060218-like events off-axis.  The global rate of such events would be some 10--100 times greater than the on-axis rate, assuming wide opening angles in the range of $\sim$10--30 degrees.  Such events might be uncovered by radio follow-up of Type Ic SNe.  Future survey projects such as the Large Synoptic Survey Telescope should detect more Type Ic SNe with a double-peaked signature of cooling envelope emission, expanding the number of potential interesting targets for radio follow-up. 

Whether or not the long prompt emission is tied to the presence of a circumstellar envelope is an interesting open question.  Clearly, this is so for the model of \citet{nakar15}.  For our model, though, the prompt X-ray and optical emission have different origins, so it may be possible to observe an X-ray signal akin to that of GRB 060218 with no prompt optical counterpart.  (On the other hand, if the envelope plays a crucial role in jet dissipation, it may still be needed.)  The high-$T_{90}$, high-variability light curves of ultra-long GRBs do seem to suggest the possibility of intrinsically long-lasting jet emission.   Interestingly, several ultra-long bursts (e.g., GRB 101225A, GRB 111209A, and GRB 121027A) also show an early optical peak that may be consistent with shock cooling \citep{levan13}.  In other ultra-long GRBs (e.g., GRB 130925A and GRB 090417B), no optical light was detected, but the presence of early optical emission cannot be ruled out due to the high extinction to those events \citep{holland10,evans14}. Overall, prompt optical emission is observed more often than not in very long bursts, hinting at one of two possibilities: either the engine duration is long because a circumstellar envelope is present, or a circumstellar envelope is present because the progenitors of long-duration engines also tend to have circumstellar envelopes.  This topic is of considerable theoretical interest going forward.  

Regardless of whether objects like GRB 060218 are powered by a jet or a shocked envelope, circumstellar interaction clearly has a role to play in explaining these unusual LLGRBs.  Both our model and the \citet{nakar15} model can be taken as further indirect evidence for the existence of a dense environment immediately surrounding the progenitor star, indicative of strong pre-explosion mass loss or binary evolution.  The mechanism driving this mass loss is unclear, but possibilities include late unstable nuclear burning \citep{sa14}, gravity wave-driven mass loss \citep{qs12}, or common envelope evolution \citep{thone11,chevalier12}.  Alternately, the circumstellar envelope could arise from a stripped binary scenario as in Type IIb SNe.  We emphasize that the progenitor's pre-explosion history is a crucial factor in determining the observed radiation's characteristics, and that this theme applies broadly to many transients including Type Ia-CSM, Type IIn, and Type IIb SNe.  Understanding the late phases of intermediate- to high-mass stellar evolution will play a critical role as our ability to detect transient phenomena continues to evolve.

%---------------------------------------------------------------------------------------------------------------------------
%***********************************************SECTION 6**************************************************
%---------------------------------------------------------------------------------------------------------------------------

\section{Conclusions}
\label{conclusions}

 We have presented a comprehensive model for the unique LLGRB GRB 060218 that provides reasonable explanations for each of its features.  The model includes a peculiar engine-driven jet with a low luminosity ($L_{iso} \sim 3 \times 10^{49} \text{ erg s}^{-1}$) and a long duration ($t_j \sim 3000$ s), properties that we suggest are related to a non-standard progenitor.  We have shown that, if the jet dissipates some modest fraction of its kinetic energy into thermal radiation, Comptonization of seed thermal photons by hot electrons can explain features of the prompt spectrum, light curve, and peak energy evolution.  We investigated different emission sites for the IC process, and found that scattering from electrons in the reverse-shocked gas can roughly account for the prompt X-ray light curve and peak energy decay, if the fraction of energy put into magnetic fields and into electrons in the forward-shocked gas is small.   Scattering from a nonthermal electron population within a dissipative jet outflow also remains a possibility for the prompt emission.  Scattering from forward shock electrons can be ruled out, as the light curves and peak energy cannot be reproduced in this case.  We also argued against a synchrotron origin for the prompt emission.  

We analysed constraints on the jet properties from the prompt thermal emission, the radio afterglow, and dynamical considerations.  There exists a region of parameter space that can fit both the radio afterglow and the prompt thermal emission without violating other constraints, although there is considerable degeneracy that prevents precise determination of the parameters.    The early thermal emission and the late-time radio afterglow can be explained either by a cold jet with relatively high energy and Lorentz factor and relatively little postshock energy in electrons and magnetic fields, or by a hot jet with lower energy and Lorentz factor and standard choices for $\epsilon_e$ and $\epsilon_B$.  Our IC model for the prompt emission breaks this degeneracy, strongly preferring the former scenario.  We derived the jet parameters $E_k \simeq 4 \times 10^{50}$\,ergs, $\gamma \simeq 11$, and $\theta_0 \simeq 0.1$, and find that the immediate circumstellar environment has a density profile of $\rho_1 \propto r^{-2.7}$ and wind parameter $A_* \simeq 4900$.  The inferred microphysical parameters of the reverse shock are $\epsilon_{e3} \simeq 6 \times 10^{-3}$ and $p=3.8$.  Combining the radio and prompt X-ray models, we constrained the magnetic parameter to $10^{-5.5} \la \epsilon_B \la 10^{-4}$ and the electron energy fraction in the forward shock to $10^{-5.5} \la \epsilon_{e2} \la 10^{-3.5}$.  Radio observations constrain the density at $r > 10^{16}$\,cm to be constant, with $n \sim 10^{3.5}$--$10^{5.5}$\,cm$^{-3}$ depending on the values of $\epsilon_{e2}$ and $\epsilon_{B}$. However, there is some concern that the outflow will not have time to sphericize prior to the radio observations, which makes the shallow radio light curve difficult to interpret.  Our results suggest that GRB 060218 may be an engine-driven event that has the same kinetic energy coupled to relativistic ejecta as in typical GRBs, but radiates very inefficiently in comparison.  This result has interesting implications considering the high volumetric rate of LLGRBs.
 
We have shown as well that the early peak in optical/UV can be powered by interaction of the fast outer SN layers with a low-mass extended envelope surrounding the progenitor star.  With the SN parameters inferred for SN 2006aj, and the luminosity and blackbody radius implied by the measured host extinction, we derive the envelope parameters $M_{ext} \approx 4 \times 10^{-3} M_\odot$ and $R_{ext} \approx 9 \times 10^{12}$ cm.  SN 2006aj is perhaps the best case so far of a double-peaked light curve characterized by cooling envelope emission, as described by \citet{np14}.  

We also tested the idea that the unusual X-ray afterglow in GRB 060218 is a dust echo of the prompt emission, as suggested by \citet{margutti14} to explain the extremely soft afterglow spectrum.  Using the available prompt emission data as an input, we modeled the expected dust echo emission from a shell of dust at $R_d$, with scattering optical depth $\tau_d$ at 1 keV.  Assuming dust grains distributed uniformly in size from a minimum radius $a_- = 0.005 \text{ }\mu\text{m}$ to a maximum radius $a_+ = 0.25 \text{ }\mu\text{m}$, we found that $\tau_d = 0.006$ and $R_d \simeq 35$ pc gave a good fit to the afterglow light curve and the spectral index evolution.  Because the echo emission does not depend on the prompt emission mechanism, this result is robust, making GRB 060218 quite a convincing case for a dust echo.  The echo model implies only a modest amount of dust consistent with the dust content of the ISM.    That the dust echo dominates over the usual synchrotron afterglow can be explained in this case by a low value of $\epsilon_e$ and $\epsilon_B$, consistent with our radio estimates and prompt X-ray modeling.  We compared our results for GRB 060218 to the other bursts with soft afterglow spectra identified by \citet{margutti14}, and found that two distinct classes of echo-dominated afterglows are indicated: one requiring a typical amount of dust (like GRB 060218), and one requiring an unusually high amount of dust (like GRB 130925A).  

We conclude by noting that our understanding of the class of low-luminosity, ultra-long GRBs with smooth light curves is severely hindered by the the small sample size -- presently, GRB 060218 and GRB 100316D are the only constituent members of this class.  In addition, because GRB 100316D lacks a detection of prompt optical emission or clear-cut evidence for prompt thermal emission, we are unable to draw any firm conclusions about it in our model. More observations of this unique class of objects is needed to settle questions about the prompt emission mechanism and the transition from beamed to spherical outflow, to better constrain the properties of the progenitor, envelope, jet, and CSM.  With the rates estimated by \citet{soderberg06}, \textit{Swift} should turn up a new burst in this class every several years or so.  In the meantime, advancing our theoretical understanding of shock-envelope interaction, the emission mechanism in relativistic jets, and the propagation of jets in complex circumstellar environments can furnish testable predictions for the next observed event.

%---------------------------------------------------------------------------------------------------------------------------
%*********************************************ACKNOWLEDGEMENTS*************************************
%---------------------------------------------------------------------------------------------------------------------------

\section*{Acknowledgements}

We thank R. Barniol-Duran and B. Morsony for helpful discussions.  This research was supported in part by NASA Grant NNX12AF90G.  

%%%%%%%%%%%%%%%%%%%%%%%%%%%%%%%%%%%%%%%%%%%%%%%%%%

%%%%%%%%%%%%%%%%%%%% REFERENCES %%%%%%%%%%%%%%%%%%

%%%%%%%%%%%%%%%%%%%%%%%%%%%%%%%%%%%%%%%%%%%%%%%%%%

%%%%%%%%%%%%%%%%% APPENDICES %%%%%%%%%%%%%%%%%%%%%

\appendix
\onecolumn

%---------------------------------------------------------------------------------------------------------------------------
%*******************************************APPENDIX A**************************************************
%---------------------------------------------------------------------------------------------------------------------------

\section{Shock dynamics of a relativistic outflow interacting with a power-law CSM}
\label{appendix_a}

Analytical solutions of equations (\ref{fdefine})$-$(\ref{rho4}) are available when $\gamma \gg 1$.   We consider the general case where the outer density profile is a power law in radius as in equation (\ref{rho4}), and the luminosity of thermal photons and the kinetic luminosity of the jet vary as power laws in time, i.e. $L_{th} =L_0 (t/t_L)^k$ and $L_{iso} =\mathcal{L}_0 (t/t_M)^s$.  We therefore have $M_{iso}(t_{emit}) \approx L_{iso} \gamma^{-1} c^{-2} (t_{emit}/t)^s$.  Combining equations (\ref{fdefine}), (\ref{rho1}), and (\ref{rho4}) with the above expressions leads to 
\begin{equation}
\label{f}
f = C_0 A_*^{-1} L_{iso,48} \gamma^{-2} \left(\dfrac{t_{emit}}{t} \right)^s \left( \dfrac{R}{R_{ext}} \right)^{\alpha-2} \left( \dfrac{R_{fs}}{R_{rs} }\right)^2.
\end{equation}
$C_0 = 5.9 \times 10^3$ is a dimensionless constant determined by scaling the density to $A_*$ and $L_{iso}$ to $10^{48}$ ergs s$^{-1}$.  Note that, in our model, $L_{th}$ and $L_{iso}$ are related by equation (\ref{Lj}); when $\gamma \gg 1$ we have
\begin{equation}
\label{Liso_48}
L_{iso,48} \approx 2.3 \times 10^{-3} \xi^{-2} L_{th,46}^{1/2} \gamma^4.
\end{equation}
For the sake of convenience and generality we do not make this substitution yet.

Three dynamical limits are possible, depending on the relative value of $f$ and $\gamma$ \citep{sp95}:
\begin{enumerate}
\item \textbf{The coasting regime ($f  \gg \gamma^2$)}:  The FS coasts with an approximately constant Lorentz factor, and the RS is Newtonian with $\bar{\beta}_3 \ll 1$:
\begin{equation}
\label{coasting}
\begin{array}{lr}
\gamma_2 \approx \gamma \\
\bar{\beta}_3 \approx \left( \dfrac{8\gamma^2}{7f}\right)^{1/2} \\
R = \dfrac{\beta_2 ct}{1-\beta_2} \approx 2 \gamma^2 ct \\
t_{emit} = \dfrac{\beta-\beta_2}{\beta(1-\beta_2)} t \approx 2 \bar{\beta}_3 t .
\end{array}
\end{equation}
The shocked regions are thin; the forward shock's size is $\sim R/\gamma^2$, and the reverse shock is even thinner by a factor $\bar{\beta_3}$, so that $R_{rs} \approx R \approx R_{fs}$ is a good approximation.  
\item \textbf{The decelerating (or accelerating) regime ($\gamma^{-2} \ll f \ll \gamma^{2}$}): The FS and the RS are both relativistic:
\begin{equation}
\label{accel}
\begin{array}{lr}
 \gamma_2 \approx \left(\dfrac{f \gamma^2}{4}\right)^{1/4} \\
\bar{\gamma}_3 \approx \left(\dfrac{\gamma^2}{4f}\right)^{1/4} \\
R \approx 2 \gamma_2^2 ct \\
t_{emit} \approx t.
\end{array}
\end{equation}
Accelerating or decelerating cases are possible, depending on the evolution of $f$.  As in the coasting case, the shocked regions are thin compared to their radius, so that $R_{rs} \approx R \approx R_{fs}$ applies.
\item \textbf{The nonrelativistic regime ($f \ll \gamma^{-2}$)}: A third solution is also possible in which the FS becomes nonrelativistic and $R_{rs} \ll R_{fs}$.  As the requisite high CSM density and low engine Lorentz factor are unlikely to be encountered in GRB 060218, we do not discuss this scenario further.
\end{enumerate}

The CSM density $\rho_1 \propto R^{-\alpha} \propto (\gamma_2^2 t)^{-\alpha}$ depends on time implicitly through $\gamma_2$. It is useful to separate out the explicit time dependence by defining $B_* = A_* (t/t_{ext})^{2-\alpha}$, with $t_{ext} \equiv R_{ext}/c$.  Additionally, we make the convenient definition $\ell = 2-\alpha$.  Then, by substituting equation (\ref{f}) into equation (\ref{coasting}) or (\ref{accel}), one obtains solutions for the dynamical variables after some algebra:
\begin{equation}
\label{fdyn}
	f = \left\{
		\begin{array}{lr}
		\left[C_1 L_{iso,48} B_*^{-1} \gamma^{-(\ell+2)+(s-\ell)} 
		\right]^{2/(s+2)},
		 & f > \gamma^2 \\
		\left[C_0 L_{iso,48} B_*^{-1} \gamma^{-(\ell+2)} 
		\right]^{2/(\ell+2)},
		 &  \gamma^{-2} < f < \gamma^2 \\
		\end{array}
	\right. ,
\end{equation}
\begin{equation}
\label{gamma2dyn}
	\gamma_2 = \left\{
		\begin{array}{lr}
		\gamma,
		 & f > \gamma^2 \\
		2^{-1/2} \left[ C_0 L_{iso,48} B_*^{-1} \right]^{1/2(\ell+2)}, 
		 &  \gamma^{-2} < f < \gamma^2 \\
		\end{array}
	\right. ,
\end{equation}
and
\begin{equation}
\label{gamma3dyn}
	\bar{\beta}_3 \bar{\gamma}_3 = \left\{
		\begin{array}{lr}
		\left(\frac{8}{7}\right)^{1/2} \left[C_1 L_{iso,48} B_*^{-1} \gamma^{-2(\ell+2)} \right]^{-1/(s+2)},
		 & f > \gamma^2 \\
		2^{-1/2} \left[C_0 L_{iso,48} B_*^{-1} \gamma^{-2(\ell+2)}\right]^{-1/2(\ell+2)}, 
		 &  \gamma^{-2} < f < \gamma^2 \\
		\end{array}
	\right. .
\end{equation}
We have defined $C_1 = (32/7)^{s/2} 2^{-\ell} C_0$, as this quantity appears repeatedly.  We see that the essential dynamical variables all depend on $B_*^{-1} L_{iso} \propto t^{s-\ell}$, allowing for two possibilities.  If $s < \ell$ we have the typical case where the forward shock begins in a coasting state and starts to decelerate once $f \sim \gamma^2$.  If $s > \ell$, however, the forward shock starts with $\gamma_2 < \gamma$ and accelerates until it reaches a terminal Lorentz factor $\gamma$ when $f \sim \gamma^2$.  When $s=\ell$, the shock velocity is constant in time; this special case generalizes the result of \citet{ec87}, who studied a constant luminosity outflow ($s=0$) in a wind density profile ($\ell=0$).  The usual afterglow dynamics \citep[e.g.,][]{sp95,spn98} can be recreated with $s=-1$ (negligible energy input) and $\ell = 2$ (constant density CSM). The transition between dynamical regimes occurs when $f \simeq \gamma^2$, at an approximate time
\begin{equation}
\label{t_f}
t_f \simeq 
\left[C_0^{-1} \mathcal{L}_{0,48}^{-1} A_* \gamma^{2(\ell+2)} t_{ext}^{-\ell} t_M^s
\right]^{1/(s-\ell)},
\end{equation}
although the exact time of transition differs for $f$, $\gamma_2$, and $\bar{\gamma}_3$ due to different leading numerical factors.

To obtain the spectral parameters in Appendix \ref{appendix_b}, it is useful to have expressions for $N_2$ and $N_3$, the number of electrons contained in regions 2 and 3.  The number of electrons swept into the forward shock can be found by integrating over the CSM density profile: $N_2 = \chi_e \int_0^R 4 \pi r^2 \rho_1 dr/m_p $, where $m_p$ is the proton mass  and $\chi_e$ is the average number of electrons per nucleon.  We take $\chi_e = 0.5$, appropriate for hydrogen-free gas, which leads to
\begin{equation}
\label{N2}
	N_2 = \dfrac{5.6 \times 10^{49}}{ (\ell+1)}  \times \left\{
		\begin{array}{lr}
		2^{\ell+1} B_* \gamma^{2(\ell+1)} t_3,
		 & f > \gamma^2 \\
		\left[C_0^{\ell+1} L_{iso,48}^{\ell+1} B_*  \right]^{1/(\ell+2)} t_3, 
		 &  \gamma^{-2} < f < \gamma^2 \\
		\end{array}
	\right. .
\end{equation}
$t_3$ is the time in units of $10^3$ s.  The number of electrons in the reverse-shocked region is $N_3 = \chi_e \int_0^{t_{emit}} L_{iso} \gamma^{-1} dt/m_p c^2$, which gives 
\begin{equation}
\label{N3}
	N_3 = \dfrac{5.6 \times 10^{49}}{(s+1)} \times \left\{
		\begin{array}{lr}
		2^\ell \left(\frac{32}{7}\right)^{1/2} \left[C_1 L_{iso,48} B_*^{s+1} \gamma^{2(\ell+2)(s+1)-(s+2)}
		\right]^{1/(s+2)} t_3,
		 & f > \gamma^2 \\
		C_0 L_{iso,48} \gamma^{-1} t_3, 
		 &  \gamma^{-2} < f < \gamma^2 \\
		\end{array}
	\right. .
\end{equation}
Note that $N_3 \sim f^{1/2} N_2$.  The comoving optical depth of region 2 can be estimated as $\tau_2 \sim (\sigma_T N_2)/(4 \pi R^2)$, and likewise for region 3.

%---------------------------------------------------------------------------------------------------------------------------
%*******************************************APPENDIX B**************************************************
%---------------------------------------------------------------------------------------------------------------------------

\section{Inverse Compton radiation from a relativistic outflow interacting with a power-law CSM}
\label{appendix_b}

The FS and RS, if at least mildly relativistic, will produce relativistic electrons and strong magnetic fields that give rise to nonthermal emission.  We adopt the standard theory, wherein fractions $\epsilon_e$ and $\epsilon_B$ of the total postshock energy density go into relativistic electrons and magnetic fields, respectively.  The postshock electron energies are assumed to be distributed as a power law, $N_{\gamma_e} \propto (\gamma_e-1)^{-p}$, above some minimum Lorentz factor $\gamma_m$.  We have $\gamma_{m2} = 610 \epsilon_{e2} g_p (\gamma_2-1)$ and $\gamma_{m3} = 610 \epsilon_{e3} g_p (\bar{\gamma}_3-1)$ for regions 2 and 3, respectively \citep[e.g.,][]{spn98}.  $g_p = 3(p-2)/(p-1)$ scales the results to $p=2.5$.

Let $P(\gamma_e)$ be the power radiated by a relativistic electron, and $\nu(\gamma_e)$ be the frequency of that radiation.  We have $P(\gamma_e) = (4/3) \sigma_T c \gamma_e^2 \gamma_2^2 u_{rad} $,where $\sigma_T$ is the Thomson cross section and  $u_{rad} = L_{th}/(4 \pi R^2 \gamma_2^2 c)$ is the photon energy density in the comoving frame.  An electron with $\gamma_e$ emits at frequency $\nu(\gamma_e) = \gamma_2 \gamma_e^2 \nu_{rad} $, where $\nu_{rad} \sim k_B T_0/h\gamma_2$ is the frequency of a typical thermal photon in the shock frame \citep{rl}. Electrons above the critical Lorentz factor $\gamma_c = (3 m_e c)/(4 \sigma_T  \gamma_2 u_{rad} t)$ can cool in time $t$ \citep{spn98, dai06}.  $\gamma_c $ is the same for regions 2 and 3 because the energy density and bulk Lorentz factor are equal across the contact discontinuity.  In the single scattering limit, the maximum spectral power emitted by an ensemble of $N_e$ electrons will be $L_{\nu,max} \approx N_e P(\gamma_e)/\nu(\gamma_e)$ \citep{spn98,dai06}.  Ignoring the self-absorption frequency $\nu_a$, which falls well below the X-ray band, the spectrum will have two breaks, at $\nu_c = \nu(\gamma_c)$ and $\nu_m = \nu(\gamma_m)$, and two possible shapes depending on whether $\nu_m < \nu_c$ (slow cooling) or $\nu_c < \nu_m$ (fast cooling).  The form of $L_\nu$ in either case is given by equations (7) and (8) in \citet{spn98}.

If the above expressions give $\gamma_{m2} < 1$, $\gamma_{m2} \approx 1$ should be used. However, in that case, only a fraction $N_{rel}/N_2 \approx [610\epsilon_{e2} g_p (\gamma_2-1)]^{p-1}$ of the electrons are relativistic with $\gamma_e -1 \ge 1$.  We make the approximation that nonrelativistic electrons do not contribute significantly to the emission at $\nu > kT_0$.  Then the spectrum above $\nu_{m2}$, where $L_\nu \propto L_{\nu,max2} \nu_{m2}^{(p-1)/2} \propto N_{rel} \gamma_{m2}^{p-1}$, is unchanged whether $\gamma_{m2} > 1$ or $\gamma_{m2} \approx 1$.  The same applies for $\gamma_{m3}$.

With the above assumptions and the dynamical equations of Appendix \ref{appendix_a}, one can compute the spectral parameters. For IC, the characteristic frequencies in region 2 and 3 are
\begin{equation}
\label{num2_ic}
	h \nu_{m2} = 630 \text{ keV} \times \left\{
		\begin{array}{lr}
		g_p^2 \epsilon_{e2,-1}^2 \xi \gamma^2,
		 & f > \gamma^2 \\
		\frac{1}{2} g_p^2 \epsilon_{e2,-1}^2 \xi \left[C_0 L_{iso,48} B_*^{-1} \right]^{1/(\ell+2)}, 
		 &  \gamma^{-2} < f < \gamma^2 \\
		\end{array}
	\right. .
\end{equation}
and
\begin{equation}
\label{num3_ic}
	h \nu_{m3} = 630 \text{ keV} \times \left\{
		\begin{array}{lr}
		\left(\frac{4}{7}\right)^2 g_p^2 \epsilon_{e3,-1}^2 \xi \left[C_1L_{iso,48}  B_*^{-1} \gamma^{-2(\ell+2)}
		\right]^{-4/(s+2)},
		 & f > \gamma^2 \\
		\frac{1}{2} g_p^2 \epsilon_{e3,-1}^2 \xi \left[ C_0 L_{iso,48} B_*^{-1} \gamma^{-2(\ell+2)}\right]^{-1/(\ell+2)}, 
		&  \gamma^{-2} < f < \gamma^2 \\
		\end{array}
	\right. .
\end{equation}
Assuming IC is the dominant cooling process, the cooling frequency is
\begin{equation}
\label{nuc_ic}
	h \nu_c = 3.0 \times 10^{-6} \text{ keV} \times \left\{
		\begin{array}{lr}
		L_{th,46}^{-2} \xi \gamma^{10} t_3^2,
		 & f > \gamma^2 \\
		\frac{1}{32} L_{th,46}^{-2} \xi \left[C_0 L_{iso,48} B_*^{-1} \right]^{5/(\ell+2)} t_3^2, 
		 &  \gamma^{-2} < f < \gamma^2 \\
		\end{array}
	\right. .
\end{equation}
The peak IC spectral power in region 2 (in cgs units) is
\begin{equation}
\label{Lnu2_ic}
	L_{\nu,max2} = \dfrac{1.1 \times 10^{27}}{ (\ell+1)} \times \left\{
		\begin{array}{lr}
		2^{\ell-1} L_{th,46} \xi^{-1} B_* \gamma^{2(\ell-1)} t_3^{-1},
		 & f > \gamma^2 \\
		L_{th,46} \xi^{-1} \left[C_0^{\ell-1} L_{iso,48}^{\ell-1} B_*^3 
		\right]^{1/(\ell+2)} t_3^{-1},
		 &  \gamma^{-2} < f < \gamma^2 \\
		\end{array}
	\right. .
\end{equation}
and in region 3 it is
\begin{equation}
\label{Lnu3_ic}
	L_{\nu,max3} = \dfrac{1.1 \times 10^{27}} {(s+1) } \times \left\{
		\begin{array}{lr}
		2^{\ell-2} \left(\frac{32}{7}\right)^{1/2} L_{th,46} \xi^{-1} 
		\left[C_1 L_{iso,48} B_*^{s+1} \gamma^{2(\ell+2)(s+1)-5(s+2)}
		\right]^{1/(s+2)} t_3^{-1},
		 & f > \gamma^2 \\
		L_{th,46} \xi^{-1} \left[C_0^{\ell} L_{iso,48}^{\ell} B_*^2 
		\gamma^{-(\ell+2)}  \right]^{1/(\ell+2)} t_3^{-1},
		 &  \gamma^{-2} < f < \gamma^2 \\
		\end{array}
	\right. .
\end{equation}
Table \ref{ICtable} summarizes the time behaviour of the spectral parameters.  We point out that, for $s > \ell$, $\nu_m$ tends towards a steep decay $\propto t^{-4(s-\ell)/(s+2)}$ as the coasting regime is approached, making this model particularly well-suited to describing GRB 060218 or other objects where a rapid decline in $E_p$ is observed.

\begin{table}
\begin{tabular}{| l | c | c |}
\hline
Quantity & Coasting ($f > \gamma^2$) & Accelerating ($f < \gamma^2$) \\
\hline
$\nu_{m2}$ & 0 & $\frac{s-\ell}{\ell+2}$ \\
\hline
$\nu_{m3}$ & $\frac{-4(s-\ell)}{s+2}$ & $\frac{-(s-\ell)}{\ell+2}$ \\
\hline
$\nu_c$ & $-2(k-1)$ & $-2(k-1) + \frac{5(s-\ell)}{\ell+2}$ \\
\hline
$L_{\nu,max2}$ & $(k-1)+\ell$ & $(k-1)+s -\frac{3(s-\ell)}{\ell+2}$ \\
\hline
$L_{\nu,max3}$ & $(k-1) + \ell + \frac{s-\ell}{s+2}$ & $(k-1)+s-\frac{2(s-\ell)}{\ell+2}$ \\
\hline
\end{tabular}
\caption{Temporal evolution of the spectral parameters.  Each parameter in the leftmost column evolves as a power-law in time, with the power-law index in the coasting and accelerating regimes given in the centre and right columns, respectively.}
\label{ICtable}
\end{table}

Given equations (\ref{num2_ic})--(\ref{Lnu3_ic}), the spectrum of the forward (reverse) shock emission can be constructed for any ordering of $\nu_c$, $\nu_{m2}$ ($\nu_{m3}$), and the observed frequency $\nu$ according to \citet{spn98}.  If the flux density at $\nu$ and the peak energy are measured at $t \ll t_f$ or $t \gg t_f$, and the relationship between $\nu$, $\nu_c$, and $\nu_m$ is known, then we can invert the model expressions for flux and peak energy to solve for two of the parameters $\gamma$, $A_*$, and $\epsilon_e$ in terms of observables and the third parameter.  As a practical example, consider the case where the reverse shock dominates the emission at $t \ll t_f$ and $\nu_c < \nu < \nu_{m3}$.  In this limit we have $L_\nu = L_{\nu,max3} \nu_c^{1/2} \nu^{-1/2}$ and $E_p = \nu_{m3}$;  $\nu_{m3}$, $\nu_c$ and $L_{\nu,max3}$ are given by equations (\ref{num3_ic}), (\ref{nuc_ic}), and (\ref{Lnu3_ic}) respectively, taking the solution for $f < \gamma^2$.  The luminosity integrated over a frequency range $\nu_1$--$\nu_2$ (with $\nu_c < \nu_1$ and $\nu_2 < \nu_{m3}$) is
\begin{equation}
\label{lightcurve}
L_{int} = 8.1 \times 10^{40} \text{ erg s}^{-1} (s+1)^{-1} C_{int} \xi^{-1/2} \left[C_1^{2\ell+5} L_{iso,48}^{2\ell+5} B_*^{-1} 
\right]^{1/2(\ell+2)} \gamma^{-1},
\end{equation}
where $C_{int} = 2\left[(h\nu_2/\text{keV})-(h\nu_1/\text{keV})\right]$.  Defining two dimensionless quantities that depend on the observables, $\tilde{L}_{int} = L_{int}/(8.1 \times 10^{40} \text{\,erg\,s}^{-1})$ and $\tilde{E}_p = E_p/(7.6 \times 10^{19} \text{\,Hz})$, considerably simplifies the algebra.  Equations (\ref{num3_ic}) and (\ref{lightcurve}) can be rewritten to eliminate either $B_*$ or $L_{iso}$:
\begin{equation}
\label{simplify1}
\tilde{L}_{int}^2 \tilde{E}_p = (s+1)^{-2} C_{int}^2 \left(g_p \epsilon_{e3,-1}\right)^2 \left(C_0 L_{iso,48}\right)^2
\end{equation} 
and
\begin{equation}
\label{simplify2}
\tilde{L}_{int} \tilde{E}_p^{(2\ell+5)/2} =  (s+1)^{-1} C_{int} \left(g_p \epsilon_{e3,-1}\right)^{2\ell+5} \xi^{\ell+2}
B_* \gamma^{2(\ell+2)}.
\end{equation}
Finally, if the thermal photons come from a dissipative outflow as described in Section \ref{prompt_thermal}, we can substitute equation (\ref{Liso_48}) for $L_{iso,48}$ and solve for $\gamma$ and $B_*$, with the results
\begin{equation}
\label{gamma_solve}
\gamma = (13.6)^{-1/4} (s+1)^{1/4} C_{int}^{-1/4} \left(g_p \epsilon_{e3,-1}\right)^{-1/4} L_{th,46}^{-1/8} \xi^{1/2}
 \tilde{L}_{int}^{1/4} \tilde{E}_p^{1/8}
\end{equation}
and
\begin{equation}
\label{B_solve}
B_* = (13.6)^{(\ell+2)/2} (s+1)^{-\ell/2} C_{int}^{\ell/2} \left(g_p \epsilon_{e3,-1}\right)^{-(3\ell+8)/2} L_{th,46}^{(\ell+2)/4} 
\xi^{-2(\ell+2)} \tilde{L}_{int}^{-\ell/2} \tilde{E}_p^{(3\ell+8)/4}.
 \end{equation}

In GRB 060218, we have $s=k/2$ by equation (\ref{Lj}), and since $k=0.66$ \citep{liang06}, $s=0.33$.  We can estimate $\alpha$ and $p$ by assuming the coasting regime has been reached by late times.  We have $\nu_{m3} \propto t^{-4(s-\ell)/(s+2)}$ for $f > \gamma^2$, so in order to get $E_p \propto \nu_{m3} \propto t^{-1.6}$ we require $\ell=-0.6$ and $\alpha=2.6$.  This in turn gives $\nu_c \propto t^{0.68}$ and $L_{\nu,max3} \propto t^{-0.54}$ for $t \gg t_f$.  To obtain a light curve $L_{\nu,max3} \nu_c \nu_{m3}^{(p-1)/2} \propto t^{-2}$ at high energies, $p=3.25$ is needed, implying $g_p = 1.67$.  The corresponding high energy spectral index is $\beta_2 \approx -1.6$, consistent with the data of \citet{toma07}.  Taking the same values for $\alpha$ and $s$, we have $L_{\nu,max3} \propto t^{-1.33}$ and $\nu_c \propto t^4$ at early times when $f < \gamma^2$.  In this limit the light curve for $\nu_c < \nu < \nu_m$ goes as $L_{\nu,max3} \nu_c^{1/2} \propto t^{0.67}$, consistent with the early rise in the XRT and BAT light curves.  We conclude that $t < t_f$ at early times, and since the 0.3--10\,keV band is well below $E_p$ at early times (and presumably above $h\nu_c$), we can apply the model described above. Taking $L_{int} = L_{XRT}$, we calculate $C_{BAT} = 5.23$.  At $t=300$\,s, the 0.3--10\,keV luminosity was $L_{XRT} \approx 1 \times 10^{46}$\,erg\,s$^{-1}$ \citep{campana06} and the peak energy was $E_p \approx 23$\,keV \citep{toma07}.  Thus, we have $\tilde{L}_{int}= 1.2 \times 10^{5}$ and $ \tilde{E}_p = 7.3 \times 10^{-2}$.  Finally, we have $L_{th,46} \approx 0.2$ and $\xi \approx 1$ at 300\,s \citep{campana06}, $t_M = t_L \approx 2800$\,s \citep{campana06}, and $t_{ext} = 300$\,s (Section \ref{optical}) so $B_* \approx A_*$ at 300\,s.  Plugging all of this into equations (\ref{gamma_solve}) and (\ref{B_solve}), we find 
\begin{equation}
\label{gamma_060218}
\gamma = 5.3 \epsilon_{e3,-1}^{-1/4}
\end{equation}
and
\begin{equation}
\label{A_060218}
A_* = 0.28 \epsilon_{e3,-1}^{-(3\ell+8)/2}.
\end{equation}
When $\epsilon_{e3} = 10^{-2.5}$, we obtain $\gamma \approx 13$ and $A_* \approx 1.3 \times 10^4$, reasonably close to the results of the best-fitting numerical model.

We stress that this model is not self-consistent unless IC is more important than synchrotron, and the emission is dominated by the RS.  Here we check whether each of these conditions is satisfied.  With $\gamma$ and $A_*$ as above, we calculate $L_{iso,48} \approx 26$, $f \approx 0.36$, $\gamma_2 \approx 1.9$, $\bar{\gamma}_3 \approx 1.7$, $R \approx 6.5 \times 10^{13}$\,cm, and $\rho_1(R) \approx 4.7 \times 10^{-13}$\,g\,cm$^{-3}$ at 300\,s.  The comoving energy density of thermal photons at the shock radius is $u_{rad} = L_{th}/4\pi R^2 \gamma_2^2 c = 3.5 \times 10^{5}$\,erg\,cm$^{-3}$, while the energy density in magnetic fields is $u_B = 4 \gamma_2^2 \rho_1 c^2 \epsilon_B = 6.1 \times 10^9 \epsilon_B$\,erg\,cm$^{-3}$ \citep{spn98}.  Synchrotron is not too important if $u_B/u_{rad} \la 1$, implying $\epsilon_B \la 6 \times 10^{-5}$, similar to the numerically inferred value.

We have assumed $\nu_c < \nu < \nu_{m3}$, so that the RS luminosity is $L_{RS} = L_{\nu,max3} \nu_c^{1/2} \nu^{-1/2}$.  Whether the FS or RS dominates the emission depends on the value of $\nu_{m2}$.  If $\nu_{m2} > \nu$, then $L_{RS}/L_{FS} = (L_{\nu,max3} \nu_c^{1/2} \nu^{-1/2})/(L_{\nu,max2} \nu_c^{1/2} \nu^{-1/2}) = L_{\nu,max3}/L_{\nu,max2} \sim N_3/N_2 \sim f^{1/2}$.  This cannot be the case, since $f < 1$ at 300\,s in our model.  Instead, we require $\nu_{m2} < \nu$, so that $L_{FS} = L_{\nu,max2} \nu_c^{1/2} \nu_{m2}^{(p-1)/2} \nu^{-p/2}$, and $L_{RS}/L_{FS} \sim f^{1/2} (\nu/\nu_{m2})^{(p-1)/2}$.  Substituting $h \nu_{m2} \simeq [610 g_p \epsilon_{e2} (\gamma_2-1)]^2 kT_0$, we obtain $L_{RS}/L_{FS} \ga 1$ when $\epsilon_{e2} \la 9 \times 10^{-4} (h \nu/kT_0)^{1/2}$.  This is qualitatively similar to the numerical result in that it also suggests $\epsilon_{e2} < \epsilon_{e3}$, although the numerical model produced a tighter upper limit on $\epsilon_{e2}$.

\label{lastpage}
\end{document}